\documentclass[fleqn,usenatbib]{mnras}
\newcommand{\teff}{\mbox{$T_{\rm eff}$}}
\newcommand{\muHz}{\mbox{$\mu{\rm Hz}$}}
\newcommand{\numax}{\mbox{$\nu_{\rm max}$}}
\newcommand{\Dnu}{\mbox{$\Delta\nu$}}

\newcommand{\kep}{\mbox{\textit{Kepler}}}
\newcommand{\tess}{\mbox{\textit{TESS}}}
\newcommand{\pextreme}{\mbox{$P_{\rm extrema}$}}
\newcommand{\redbf}[1]{\textcolor{black}{\textnormal{{#1}}}}

\usepackage{tikz,xcolor,hyperref}
\definecolor{lime}{HTML}{A6CE39}
\DeclareRobustCommand{\orcidicon}{%
	\begin{tikzpicture}
	\draw[lime, fill=lime] (0,0) 
	circle [radius=0.16] 
	node[white] {{\fontfamily{qag}\selectfont \tiny ID}};
	\draw[white, fill=white] (-0.0625,0.095) 
	circle [radius=0.007];
	\end{tikzpicture}
	\hspace{-2mm}
}
\foreach \x in {A, ..., Z}{%
	\expandafter\xdef\csname orcid\x\endcsname{\noexpand\href{https://orcid.org/\csname orcidauthor\x\endcsname}{\noexpand\orcidicon}}
}
\usepackage{newtxtext,newtxmath}
\usepackage[T1]{fontenc}
\usepackage{ae,aecompl}

\usepackage{graphicx}	% Including figure files
\usepackage{amsmath}	% Advanced maths commands
\usepackage{amssymb}	% Extra maths symbols

\title[\kep\ Long Period Variables]{Asteroseismology of luminous red giants with \kep\ I:  Long Period Variables with radial and non-radial modes}

\author[Jie Yu et al.]{Jie Yu\orcidA{}$^{1,2,3}$\thanks{E-mail: yujie@mps.mpg.de (JY)},
Timothy R. Bedding\orcidB{}$^{1,3}$,
Dennis Stello\orcidC{}$^{4,1,3}$,
Daniel Huber\orcidD{}$^{5,1,3}$,
\newauthor
Douglas L. Compton\orcidE{}$^{1,3}$,
Laurent Gizon\orcidF{}$^{2,6,7}$,
and Saskia Hekker\orcidG{}$^{2,3}$
\\
\\
$^{1}$Sydney Institute for Astronomy (SIfA), School of Physics, University of 
Sydney, NSW 2006, Australia\\
$^{2}$Max Planck Institute for Solar System Research, Justus-von-Liebig-Weg 3, 37077 G\"{o}ttingen, Germany\\
$^{3}$Stellar Astrophysics Centre, Department of Physics and Astronomy, Aarhus University, Ny Munkegade 120, DK-8000 Aarhus C, Denmark\\
$^{4}$School of Physics, University of New South Wales, NSW 2052, Australia\\
$^{5}$Institute for Astronomy, University of Hawai`i, 2680 Wood-lawn Drive, Honolulu, HI 96822, USA\\
$^{6}$Institut f\"{u}r Astrophysik, Georg-August-Universit\"{a}t G\"{o}ttingen, Friedrich-Hund-Platz 1, 37077 G\"{o}ttingen, Germany\\
$^{7}$Center for Space Science, NYUAD Institute, New York University Abu Dhabi, PO Box 129188, Abu Dhabi, UAE
}
\date{Accepted XXX. Received YYY; in original form ZZZ}
\pubyear{2019}

\begin{document}
\label{firstpage}
\pagerange{\pageref{firstpage}--\pageref{lastpage}}
\maketitle

\begin{abstract}
While long period variables (LPVs) have been extensively investigated, especially with MACHO and OGLE data for the Magellanic Clouds, there still exist open questions in their pulsations regarding the excitation mechanisms, radial order  and angular degree assignment. Here, we perform asteroseismic analyses on LPVs observed by the 4-year \kep\ mission. Using a cross-correlation method, we detect unambiguous pulsation ridges associated with radial fundamental modes ($n=1$) and overtones ($n\geqslant2$), where the radial order assignment is made by using theoretical frequencies and observed frequencies. Our results confirm that the amplitude variability seen in semiregulars is consistent with oscillations being solar-like. We identify that the dipole modes, $l=1$, are dominant in the radial orders of $3\leq n \leq6$, and that quadrupole modes, $l=2$, are dominant in the first overtone $n=2$. A test of seismic scaling relations using Gaia DR2 parallaxes reveals the possibility that the relations break down when \numax~$\lesssim$~3~\muHz\ (R~$\gtrsim$~40 R$_{\odot}$, or log $\rm L/L_{\odot}$~$\gtrsim$~2.6). Our homogeneous measurements of pulsation amplitude and period for 3213 LPVs will be very valuable for probing effects of pulsation on mass loss, in particular in those stars with periods around 60 days, which has been argued as a threshold of substantial pulsation-triggered mass loss.
\end{abstract}

% We find that the amplitude of the dominant pulsation modes starts to increase more significantly with period at a period of $P=4.5$ days, which can be a result of the transition of dominant modes between overtones and may suggest significant variations in the mode lifetime.

\begin{keywords}
stars: oscillations, stars: evolution, stars: late-type, techniques: photometric
\end{keywords}

%%%%%%%%%%%%%%%%%%%%%%%%%%%%%%%%%%%%%%%%%%%%%%%%%%

%%%%%%%%%%%%%%%%% BODY OF PAPER %%%%%%%%%%%%%%%%%%

\section{Introduction}
Long Period Variables (LPVs)\footnote{In this work, we use these three terms interchangeably: LPVs, M giants, and high-luminosity red giants, though the first is extended to include pulsators with periods down to a few days, and the second is extended to include some late K giants.} are cool evolved stars on the asymptotic giant branch or near the tip of the red giant branch. They are generally divided into Semiregular Variables (SRs) and Mira variables, based on the regularity and amplitudes of their light curves.  Major advances in the understanding of pulsations in LPVs have been achieved from studying their period--luminosity (P--L) diagrams, using ground-based surveys such as MACHO \citep{wood1999a}, EROS \citep{lebzelter2002a}, and OGLE \citep{soszynski2004a,soszynski2009a}, and space missions like \textit{Hipparcos} \citep{bedding1998b, tabur2010a}, CoRoT \citep{lebzelter2011a, ferreira-lopes2015a}, \kep\ \citep{banyai2013a,mosser2013a,hartig2014a, stello2014a}, and Gaia \citep{mowlavi2018a,lebzelter2018a}.  While the pulsation sequences of LPVs have been extensively studied, the nature of the pulsations is still not fully understood.

The first open question is linked to the driving mechanism of the pulsations in SRs: self-excitation via a heat-engine mechanism like Mira variables, or stochastically excited as solar-like oscillations in G and K stars? One method to investigate excitation mechanisms of LPVs is to analyze the relation between the pulsation amplitude and period, and compare with less-luminous red giants that are well-established to be sun-like oscillators \citep[e.g.][]{tabur2010a}. \citet{soszynski2007a} proposed that stars falling along their so-called sequences b2 and b3 are sun-like pulsators (see the sequences in Figure~\ref{fig:kperiod}a). \citet{mosser2013a} argued that all the P--L relations for LPVs can be explained by solar-like oscillations, in that the P--L sequences are an extension of a global oscillation pattern in less-evolved red giants. This is consistent with the findings by \citet{dziembowski2010a} and \citet{takayama2013a} and but discrepant to the predictions by \citet{xiong2018a}. However, \citet{banyai2013a} argued that Mira/Semiregular variables may have a  pulsation nature different from sun-like oscillations. This is based on their findings of a significant pulsation-amplitude transition at a period of $\sim$10 days, a dividing point between SRs and shorter-period solar-like pulsators. We note that \citet[][their Figure 3]{ferreira-lopes2015a} found a similar amplitude-transition feature, which, however, may be attributed to the different amplitude definitions between their work and the comparison reference. In this work, we find evidence in support of solar-like oscillations in SRs (see Section \ref{{excitation}}).

The second question concerns assigning radial orders to the pulsation sequences on the P--L diagram. \citet{soszynski2007a}, \citet{dziembowski2010a}, and \citet{mosser2013a} interpreted the sequences C$'$, B, and~A as the radial fundamental  mode, first overtone, and second overtone, respectively (see the sequences in Figure~\ref{fig:kperiod}a). This means the longer-period sequence C, containing Mira variables, has no interpretation in terms of radial orders if we assume that two adjacent sequences differ by one radial order. However, a distinct set of radial order assignment from \citet{wood1999a}, \citet{soszynski2013b}, \citet{takayama2013a}, and \citet{wood2015a} state that sequences C, C$'$, and B are associated with the radial fundamental mode, first overtone, and second overtone, respectively. Thus, the two sets of modal assignments differ by one radial order. Recently, \citet{trabucchi2017a} re-examined the observed P--L sequences and gave an intermediate solution that $\rm C^{'}$ and B both correspond to the first overtone, but include fundamental mode pulsations at lower luminosities of the two sequences. They suggested that sequences C, A, and A$'$ correspond to the radial fundamental mode, second overtone, and third overtone, respectively. We discuss this issue in Section \ref{orders}.

The third question concerns the angular degree of the modes. Do LPVs exhibit radial and non-radial pulsations? And which modes are dominant, radial modes ($l=0$), dipole modes ($l=1$), or quadrupole modes ($l=2$)? \citet{soszynski2004a} discovered that the sequence A consists of three closely separated parallel subsequences in the so-called Petersen diagram, where the ratio of a shorter period to a longer period is plotted against the longer period. The subsequences were later also found in the sequence B and A$'$ by \citet{soszynski2007a}. \citet{stello2014a} found a triplet frequency pattern in \kep\ M giants that is made up of $l=1,2,0$ modes, sorted in the decreasing period order. This triplet pattern explains the parallel subsequences in the Petersen diagram. \citet{mosser2013a} further argued that dipole modes dominate in stars oscillating at higher frequency ($\gtrsim$ 1.0 \muHz), while radial modes dominate at lower frequency ($\lesssim$ 1.0 \muHz). However, \citet{stello2014a} found that the pulsations of luminous \kep\ stars with a characteristic oscillation frequency down to 0.2 \muHz\ are dominated by dipole modes. The findings from \citet[][Figure 3]{wood2015a} using the OGLE III catalog of LPVs in the LMC is in agreement with the findings by \citet{stello2014a}. 
We address this issue in Section \ref{dominantmodes}.

The fourth question is related to the asteroseismic scaling relations. The relations have been widely used to characterize oscillating dwarfs and giants \citep[see][for reviews]{chaplin2013a, hekker2017a}. Moreover, asteroseismically derived parallaxes have been used as references to calibrate the Gaia DR2 parallaxes \citep[e.g.][]{zinn2019a}. Although the seismic scaling relations have been extensively tested on main-sequence stars, subgiants, and less-luminous red giants \citep[for a recent review, see][]{hekker2019a}, it remains an open question if the relations work for high-luminosity red giants (see Section \ref{testseis}).

\begin{figure}
\begin{center}
\resizebox{\columnwidth}{!}{\includegraphics{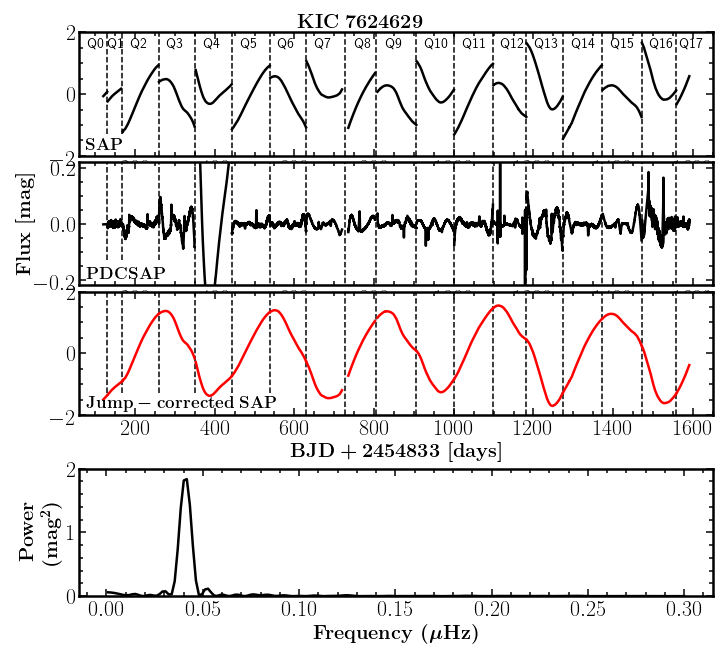}}
\resizebox{\columnwidth}{!}{\includegraphics{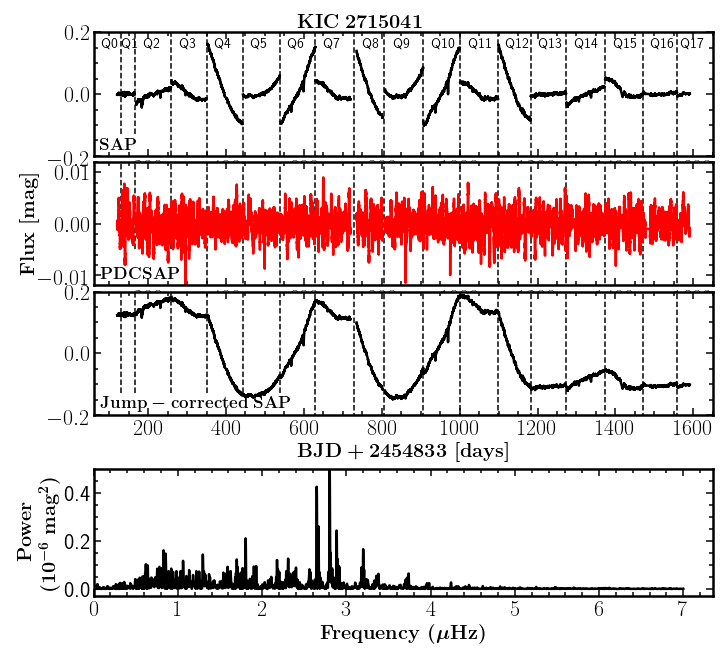}}
\caption{Light curves for two representative \kep\ LPVs: KIC~7624629 (upper, slow pulsator) and KIC~2715041(lower, fast pulsator). For each star, the SAP (1st panel), \mbox{PDCSAP} (2nd panel), and jump-corrected SAP (3rd panel) light curves are shown. The light curves in red were adopted for asteroseismic analyses. The bottom panel of each star shows the power spectrum, computed from the adopted light curves. In this work, all power spectra were calculated in units of square micro-magnitude. Our spectrum normalisation ensures that we measured the amplitude of a sinusoidal wave.}
\label{fig:lcs}
\end{center}
\end{figure}

% Jump-corrected SAP time series are adopted for slowly pulsating stars, like KIC~7624629, while PDCSAP light curves are used for fast pulsating stars, like KIC~2715041, respectively.

\begin{figure*}
\begin{center}
\resizebox{\textwidth}{!}{\includegraphics{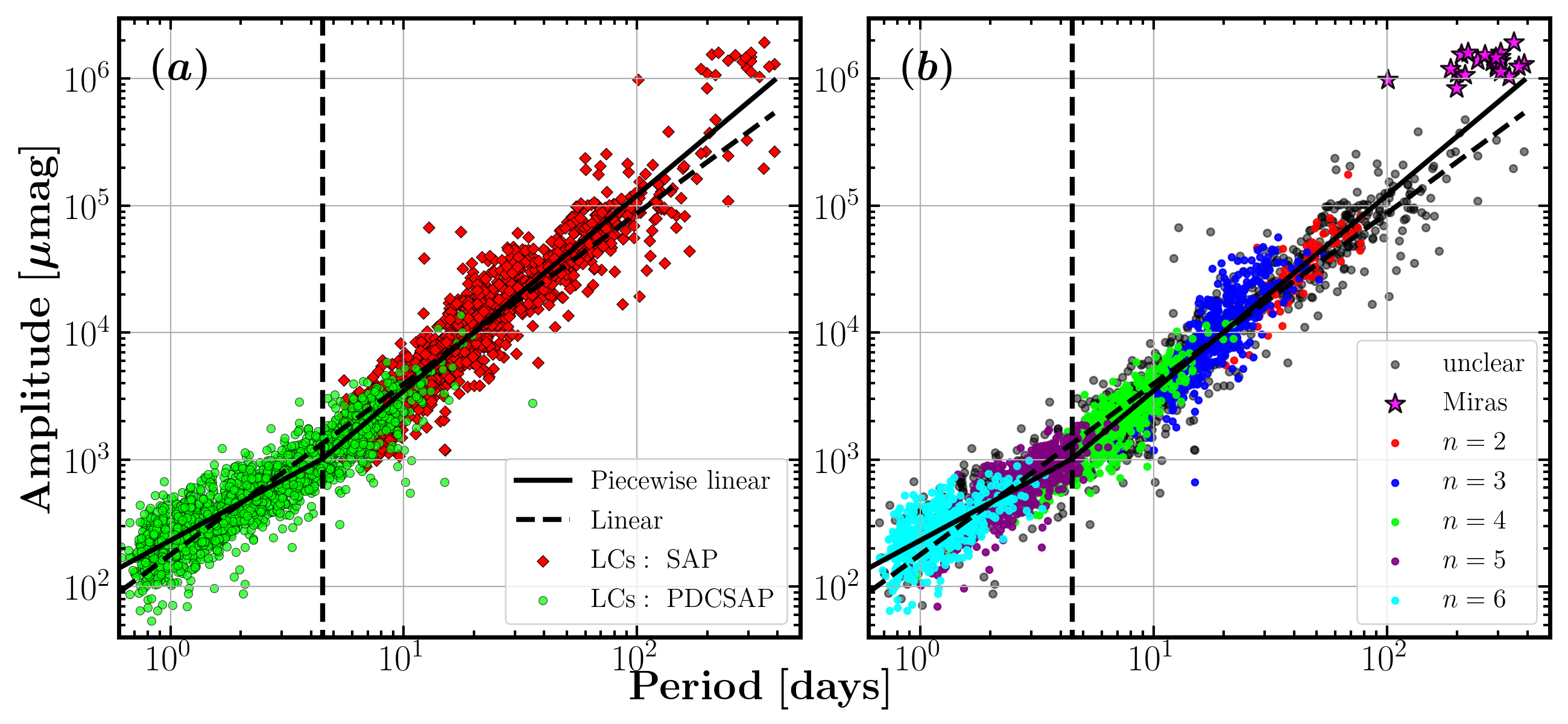}}\\
\caption{\textbf{(a)} Relation between the period and amplitude proxy of the dominant mode,  i.e., the highest peak in a power spectrum. Periods and amplitudes were extracted from either PDCSAP time series shown in green circles, or jump-corrected SAP time series indicated in red diamonds (see the text). The piecewise linear model (solid), preferred to the linear model (dashed), shows a kink at period P~$\simeq$~4.5~days, indicated in the vertical dashed line. \textbf{(b)} Similar to \textbf{(a)} now color-coded by the radial order of the dominant mode.}
\label{fig:periodamplitude}
\end{center}
\end{figure*}

In this work we address these four open questions using a sample of 3213 \kep\ LPVs, which includes pulsators with periods $P~\ga$ 1 day. Our asteroseismic analyses are based on the light curves collected by the 4-year \kep\ space mission  \citep{borucki2010a, koch2010a} and on Gaia DR2 parallaxes \citep{lindegren2018a}. 

\begin{figure*}
\begin{center}
\resizebox{0.8\textwidth}{!}{\includegraphics{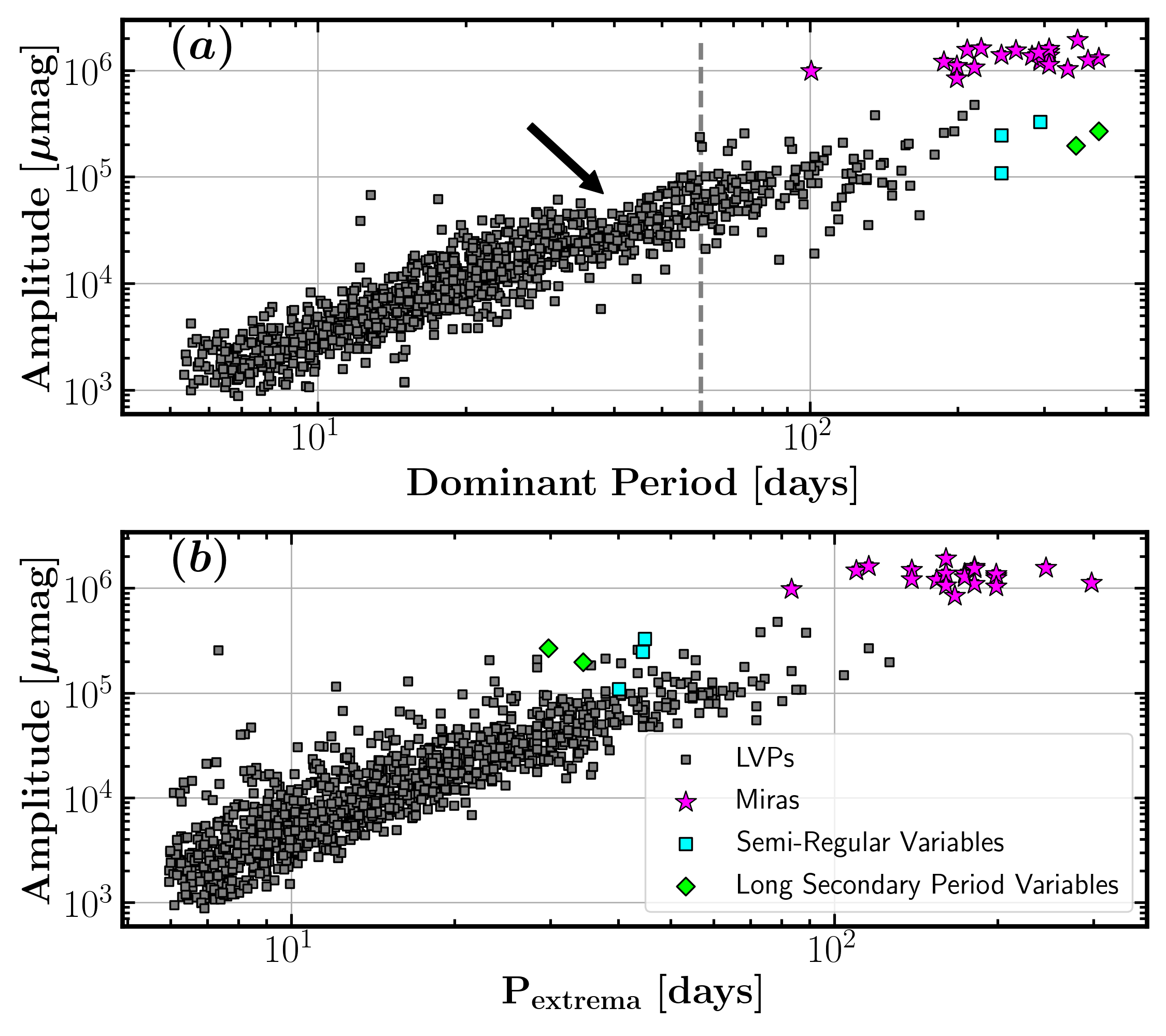}}\\
\caption{\textbf{(a)}: Relation between amplitude and period for stars with periods greater than 6 days, thus dominated by SRs and Mira variables. The dashed line marks the approximate period threshold, 60 days, above which substantial pulsation-triggered mass loss is expected. \textbf{(b)}: Relation between amplitude and \pextreme\ for the same stars. All the Long secondary period variables and Miras in the entire sample are highlighted, while only two representative SRs are shown.} 
\label{fig:zerocrossing}
\end{center}
\end{figure*}

\begin{figure*}
\begin{center}
\resizebox{\textwidth}{!}{\includegraphics{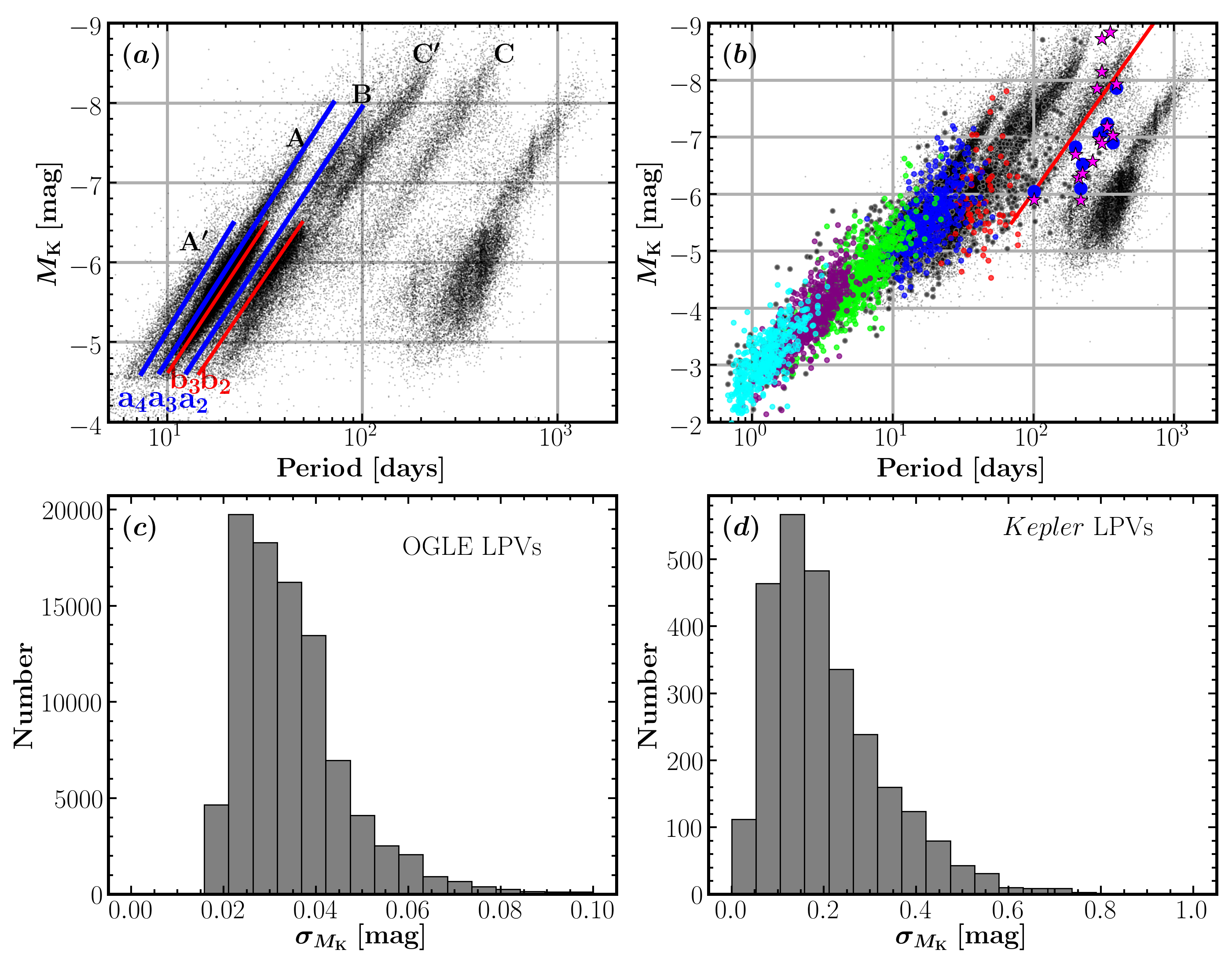}}\\
\caption{\textbf{(a)} Period-$M_K$ diagram of OGLE LPVs (black points) in the LMC \citep[dominant mode only,][]{soszynski2009a}. For OGLE small amplitude red giants, sequences $a_2$, $a_3$, and $a_4$ denoting AGB stars are shown in blue lines, while sequences $b_2$ and $b_3$ denoting red-giant-branch (RGB) stars are shown in red lines (the line parameters were adopted from Table 1 of \citealt{soszynski2007a}). \textbf{(b)} Similar to panel \textbf{a} now including the \kep\ LPVs (Note the difference in the scale of the vertical and horizontal axes). Symbol colors have the same meaning as Figure~\ref{fig:periodamplitude}b. Miras with Gaia DR2 parallaxes better than 30\% are highlighted in the dark blue circles. The red line denotes the Period-$M_K$ relation for Miras \citep{feast1996a}. \textbf{(c)} Uncertainties of $M_K$ for the OGLE LPVs. \textbf{(d)} Uncertainties of $M_K$ for the \kep\ LPVs.} 
\label{fig:kperiod}
\end{center}
\end{figure*}

%%%%%%%%%%%%%%%%%%%%%%%%%%%%%%%%%%%%%%%%%%%%%%%%%%%%%%%%%%%%%%%%%%%%%%%%%%%%%%%%
\section{Sample selection and data reduction}\label{sampledata}
To construct a sample of LPVs, we selected 4296 \kep\ red giants from \citet{mathur2017a} with surface gravity log~$\it{g}$ < 2.0 dex, equivalent to a period $\gtrsim$ 1 day. We added known \kep\ M giants from the literature, namely, \citet{banyai2013a}, \citet{stello2014a}, and \citet{yu2018a}. For the sample in \citet{yu2018a}, we applied a cut of \numax~$\leqslant~15~\muHz$. Note that the M giants collected from the literature also meet the criterion of periods $\gtrsim$ 1 day.

From the sample of 4724 stars selected above, we excluded the stars with marginal pulsation detections. This is because 75\% of them have too short light curves for our analyses, i.e., they were observed for fewer than 4 quarters, and/or too faint, the \kep\ magnitudes Kp$>$14 mag. For the other 25\% of the excluded stars (that is, 8\% stars in the original sample), no clear pulsation signal is found. It has been known that for some stars, which can be bright and have long time series, solar-like oscillations are expected but not detected \citep[e.g. see][]{chaplin2011a, mathur2019a, Schonhut-Stasik2019a}. We note that \citet{Schonhut-Stasik2019a} reported the same fraction for red giants (8\%) without solar-like oscillation detections. Our final sample comprised 3213 LPVs, as listed in Table \ref{seisparatable}. 

\kep\ long-cadence photometry is well-suited for exploring pulsations in LPVs, given its temporal sampling rate (29.4 min) and long baseline ($\sim$4 years). We used \mbox{PDCSAP} light curves, which have been corrected for systematic errors in each observing quarter using ``cotrending basis vectors" \citep{stumpe2012a, smith2012a}. For some M giants pulsating at a long period, such as Mira variables, \mbox{PDCSAP} time series were over-corrected, by treating intrinsic pulsations as ``systematic errors''. For these stars, we adopted ``Simple Aperture Photometry" (SAP) light curves. To determine the stars for which the PDCSAP light curves were safe to use, we used a measure, \pextreme, which approximates a typical period of a light curve. It is defined as 
\begin{equation}
 P_{\rm extrema} = \frac{2N\delta t}{N_{\rm extrema}},
\end{equation}
where $N_{\rm extrema}$ is the number of turning points (extrema), $N$ is the total number of data points of a light curve, and $\delta t$ is the sampling interval of the long-cadence \kep\ data (29.4 min). 

To count the turning points in a light curve, we calculated the point-to-point difference of the light curve. The number of the zero crossings of the difference time series gives the number of turning points. Since there are quarter gaps in \kep\ light curves, which will introduce turning points between quarter edges because the corresponding flux usually jumps dramatically, we then performed iterative 4-$\sigma$ clipping to discard outliers of the difference time series. Finally, we found \pextreme\ = 6 days is an appropriate threshold to select the light curve source (PDCSAP versus SAP). We have tested different thresholds of $P_{\rm extrema}$ up to 30 days, and found the conclusions below were not changed significantly. This is because our method for preparing light curves also works well in the short-period range where PDCSAP light curves were used. 

For PDCSAP time series with \pextreme < 6 days, we divided each quarter of time series by its median flux, and concatenated them together.  For SAP data, we used a Gaussian Process method \citep{rasmussen2006a} to remove jumps between two adjacent quarters for stars with \pextreme > 6 days. A step function was used to model jumps and a covariance function was used to approximate the residuals, due to actual physical brightness variations as well as noise. We implemented the Gaussian process fit using $CELERITE$ and the kernel $SHOTerm$ \citep{foreman-mackey2017a}. Note that systematic perturbations, such as spacecraft safe-mode events and/or long-term drifts, could affect light curves but were not corrected considering their much lower amplitudes than the intrinsic stellar variability. Stars for which we detected periods equal to one \kep\ orbit (372 d) were discarded. We hereafter refer to a star as a slow pulsator if \pextreme > 6 days or as a fast pulsator if \pextreme $\leqslant$ 6 days.

Figure \ref{fig:lcs} shows the jump-corrected SAP light curve (red curve) and its power spectrum, for a representative slowly pulsating M giant, KIC 7624629 (top~4 panels), with \pextreme=198.41 days.  The PDCSAP time series for this star was clearly over-corrected. Figure \ref{fig:lcs} also shows the SAP and PDCSAP light curves for a typical fast pulsating star, KIC 2715041 (bottom~4 panels), with \pextreme=3.36 days. For this star, we can see that systematic annual perturbations in the PDCSAP light curve have been nicely removed. The jump-corrected SAP light curve for this star is clearly dominated by annual instrumental drifts.

For the fast and slow samples, we used different methods to measure the pulsation period and amplitude of the dominant modes. For pulsators in the slow sample, we first measured the period of the highest peak in the power spectrum of the difference time series. Note that a difference time series is essentially free from the quarter jumps in the associated light curve, since the jumps manifest themselves as outliers that can be clipped.  In order to measure the amplitude, we then searched the power spectrum of the time series for the highest peak in a window with a width of 10 times the frequency resolution and centered at the frequency measured from the difference time series. The height of the highest peak was used as a proxy for the amplitude.

For stars in the fast sample, we first used the SYD pipeline \citep{huber2009a} to measure the frequency of maximum oscillation power, \numax. We subsequently selected the highest peak within a window centered at \numax. The width of this window was the full-width-at-half-maximum of a Gaussian fitted to the auto-correlation of the oscillation power excess. These steps are necessary, because for a fast oscillator the dominant mode generally is not the highest peak in the power spectrum, due to 1/f noise in the lower frequency regime. This is unlike a slow pulsator, such as a Mira variable, for which the highest peak generally is the dominant mode. Again, we used the height of the highest peak as a proxy to approximate the amplitude. We note that the amplitude of a dominant mode contains the contributions from the granulation background, which is the case for both fast and slow pulsators. 

To understand the impact of the background, we calculated the ratio of the height of the dominant mode to the amplitude of granulation at the frequency of the dominant mode. The granulation amplitude was calculated from the fitted background obtained when measuring the global seismic parameters, \numax\ and \Dnu, using the SYD pipeline (see Section 6.1). We found that this ratio is approximately constant, consistent with \citet[][see their Figure 4]{mosser2013a}, and has a median value of 5.9. This suggests that our measured proxy of the oscillation amplitudes globally overestimates the amplitude by $\sim$~20\% \footnote{Since (O+B)/B=5.9, i.e., O/B=4.9, one obtains H/O=(O+B)/O=1.2. Here, H, O, B denote the height of the highest peak, the oscillation amplitude proxy, and the granulation amplitude at the highest peak, respectively.}

\section{Period--amplitude relation of long period variables}
\label{lpvperiodamp}
Figure~\ref{fig:periodamplitude}a shows our measured amplitude proxy versus periods for the entire sample. At the longest periods we see a number of Mira variables with periods $P>100$ days and amplitude near 1.0 mag (also shown in the pink asterisks in Figure~\ref{fig:periodamplitude}b). We also see SRs, with periods typically longer than 20 days and lower amplitudes. Note that in this work we measured the amplitude of a sinusoidal wave, which is half of the peak-to-peak amplitude. Miras are characterized by pulsation periods longer than 100 days and peak-to-peak amplitudes greater than 2.5 mag at visual wavelengths. Here, for Mira variables a typical measured peak-to-peak amplitude is 2.0 mag, smaller than the 2.5 mag definition, which is because of the redder and broader \kep\ bandpass \citep[e.g. see][and references therein]{lund2019a}. Both period and amplitude, together with additional parameters, are listed in Table~\ref{seisparatable}.

\begin{table*}
\begin{footnotesize}
\begin{centering}
\caption{Asteroseismic parameters and stellar properties of \kep\ M giants}
\resizebox{\textwidth}{!}{\begin{tabular}{cccccccccccccc}
\hline
\hline
% \small 
   KIC&    LCs&         Q& \pextreme &        \teff &         amp&    period&  order&    \numax&                 \Dnu&             $\pi$&            $d$&                  $L$&        $A_v$\\
     -&      -&         -&       -&             K&      $\mu$mag&       day&       &     \muHz&                \muHz&             mas&             kpc&            $\rm{L_{\odot}}$&    mag\\
   (1)&    (2)&       (3)&     (4)&           (5)&           (6)&       (7)&    (8)&       (9)&                 (10)&            (11)&            (12)&                   (13)&        (14)\\
\hline

  892738& PDCSAP&      18&     nan&  4534$\pm$135&      308.24&      1.44&      6&    7.47$\pm$0.25&    1.31$\pm$0.01&   0.40$\pm$0.02&     2.490$\pm$0.137&   193.96$\pm$ 21.69&    0.29\\
  893210& PDCSAP&      17&     nan&  4204$\pm$127&     1004.15&      4.36&      5&    2.62$\pm$0.05&    0.51$\pm$0.01&   0.23$\pm$0.03&     4.440$\pm$0.503&   574.52$\pm$133.14&    0.25\\
  893233& PDCSAP&       8&     nan&  4207$\pm$147&      903.45&      1.94&      5&    6.16$\pm$0.08&    1.18$\pm$0.01&   0.43$\pm$0.03&     2.338$\pm$0.151&   227.36$\pm$ 29.88&    0.28\\
 1026309& PDCSAP&      18&     nan&  4514$\pm$80&       97.86&      0.77&      6&   16.09$\pm$0.91&    1.92$\pm$0.01&   0.67$\pm$0.02&     1.502$\pm$0.048&   178.02$\pm$ 12.19&    0.28\\
 1026895& PDCSAP&      18&     nan&  3900$\pm$80&      714.65&      3.88&      5&    2.78$\pm$0.11&    0.54$\pm$0.01&   0.83$\pm$0.02&     1.214$\pm$0.036&   445.50$\pm$ 27.16&    0.24\\
 1027110& PDCSAP&      18&     nan&  4190$\pm$80&      466.93&      1.62&      6&    6.67$\pm$0.14&    1.15$\pm$0.01&   0.33$\pm$0.02&     3.019$\pm$0.194&   187.73$\pm$ 24.44&    0.24\\
 1027707& PDCSAP&      18&     nan&  4254$\pm$148&      872.85&      3.87&      4&    3.03$\pm$0.04&    0.54$\pm$0.01&   0.17$\pm$0.03&     5.739$\pm$0.841&   654.15$\pm$198.06&    0.23\\
 1160655& PDCSAP&      18&     nan&  3740$\pm$130&     1715.10&      7.28&      4&    1.63$\pm$0.02&    0.37$\pm$0.01&   0.38$\pm$0.03&     2.631$\pm$0.181&   253.08$\pm$ 36.65&    0.61\\
 1160867& PDCSAP&      18&     nan&  4000$\pm$80&      550.43&      2.56&      4&    4.68$\pm$0.08&    0.89$\pm$0.01&   1.12$\pm$0.03&     0.893$\pm$0.022&   240.36$\pm$ 12.64&    0.70\\
 1160986& PDCSAP&       4&     nan&  4347$\pm$80&      193.40&      0.99&      7&    8.85$\pm$0.62&    1.54$\pm$0.02&   0.21$\pm$0.02&     4.688$\pm$0.334&   134.51$\pm$ 19.42&    0.71\\

12600259& PDCSAP&      18&     nan&  4288$\pm$150&      283.99&      1.38&      6&    7.69$\pm$0.2&    1.36$\pm$0.02&   0.35$\pm$0.02&     2.872$\pm$0.204&   162.86$\pm$ 23.63&    0.15\\
12600652& PDCSAP&      18&     nan&  4056$\pm$141&     1210.53&      3.98&      4&    2.86$\pm$0.04&    0.57$\pm$0.02&   0.27$\pm$0.02&     3.801$\pm$0.351&   348.10$\pm$ 65.53&    0.13\\
12601040&    SAP&      18&   32.30&  3275$\pm$114&    56380.99&     67.45&      2&          nan    &          nan    &   0.26$\pm$0.06&     3.996$\pm$0.918&  1548.26$\pm$759.13&    0.11\\
12602404& PDCSAP&      10&     nan&  4457$\pm$ 80&      315.54&      0.88&      6&   12.97$\pm$0.13&    1.95$\pm$0.01&   0.30$\pm$0.02&     3.352$\pm$0.241&   129.36$\pm$ 18.96&    0.27\\
12602421&    SAP&      18&    7.80&  4175$\pm$124&     2346.49&      7.94&    nan&          nan    &          nan    &   0.66$\pm$0.02&     1.525$\pm$0.052&   473.81$\pm$ 33.76&    0.27\\
12644223& PDCSAP&      18&     nan&  4086$\pm$143&      893.23&      2.63&      5&    4.18$\pm$0.13&    0.84$\pm$0.02&   0.16$\pm$0.01&     6.217$\pm$0.456&   281.87$\pm$ 41.99&    0.20\\
12645224&    SAP&      18&   12.66&  4035$\pm$141&     5338.09&     11.78&    nan&          nan    &          nan    &   0.40$\pm$0.03&     2.512$\pm$0.168&   857.84$\pm$116.16&    0.16\\
12688798& PDCSAP&       6&     nan&  4245$\pm$ 80&      395.85&      0.92&      6&   13.05$\pm$0.21&    1.97$\pm$0.01&   0.31$\pm$0.01&     3.262$\pm$0.150&    86.88$\pm$  8.27&    0.15\\
12690711&    SAP&      18&    6.38&  3986$\pm$139&     1422.41&      7.26&    nan&          nan    &          nan    &   0.23$\pm$0.03&     4.454$\pm$0.495&   657.43$\pm$149.19&    0.21\\
12984227&    SAP&      18&   17.36&  3489$\pm$122&     9606.24&     20.37&    nan&          nan    &          nan    &   0.28$\pm$0.04&     3.670$\pm$0.534&   880.49$\pm$264.38&    0.20\\

\hline
\end{tabular}}
\label{seisparatable}
\flushleft Note. (1) KIC ID; (2) The type of light curve used in this work; (3) Number of quarters of \kep\ light curves; (4) \pextreme\ (see Section \ref{sampledata} for its definition); (5) Source: \citet{mathur2017a}; (6) Dominant mode amplitude; (7) Dominant mode period; (8) Radial order; (9) The frequency of maximum power; (10) Mean larger frequency separation; (11) Gaia DR2 parallax with an offset of 0.03 mas added \citep{lindegren2018a}; (12) (13) (14) Distance, Luminosity, and Extinction, respectively. 
\begin{flushleft} (This table is available in its entirety in machine-readable form.)
\end{flushleft}
\end{centering}
\flushleft
\end{footnotesize}
\end{table*}

\begin{figure*}
\begin{center}
\resizebox{0.65\textwidth}{!}{\includegraphics{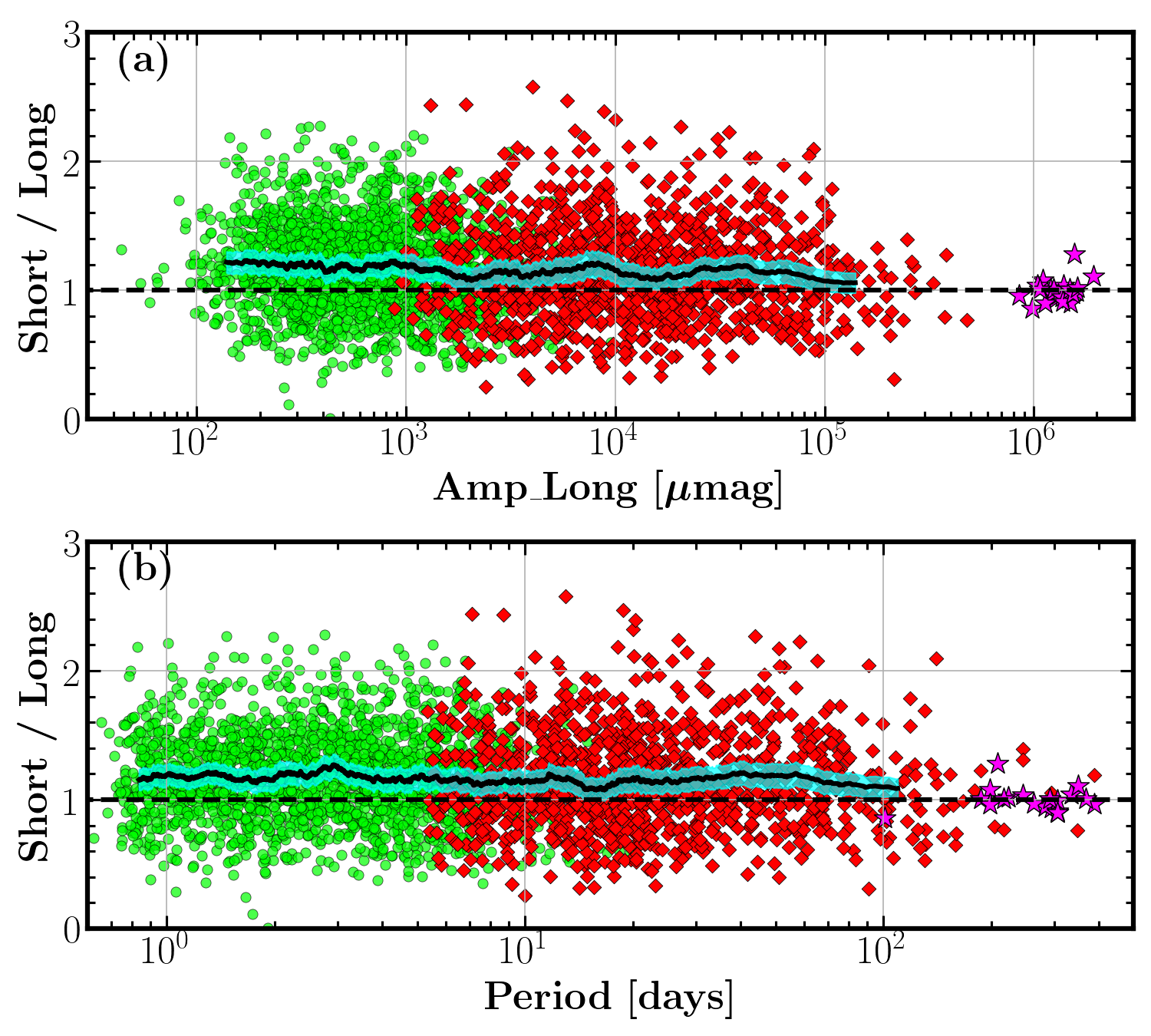}}\\
\caption{Comparison of the dominant mode amplitudes measured from full-length and a 1/3-length of the \kep\ light curves. The top panel displays the amplitude ratio as a function of the amplitude measured from full-length light curves, while the bottom panel shows the ratio against period, also determined from full-length light curves. 
The running median values are shown in black and their 3-$\sigma$ uncertainties are shown in the cyan filled region. Green and red symbols have the same meaning as Figure~\ref{fig:periodamplitude}a. Miras are highlighted with the pink asterisks.} 
\label{fig:modelifttime}
\end{center}
\end{figure*}

\begin{figure*}
\begin{center}
\resizebox{0.7\textwidth}{!}{\includegraphics{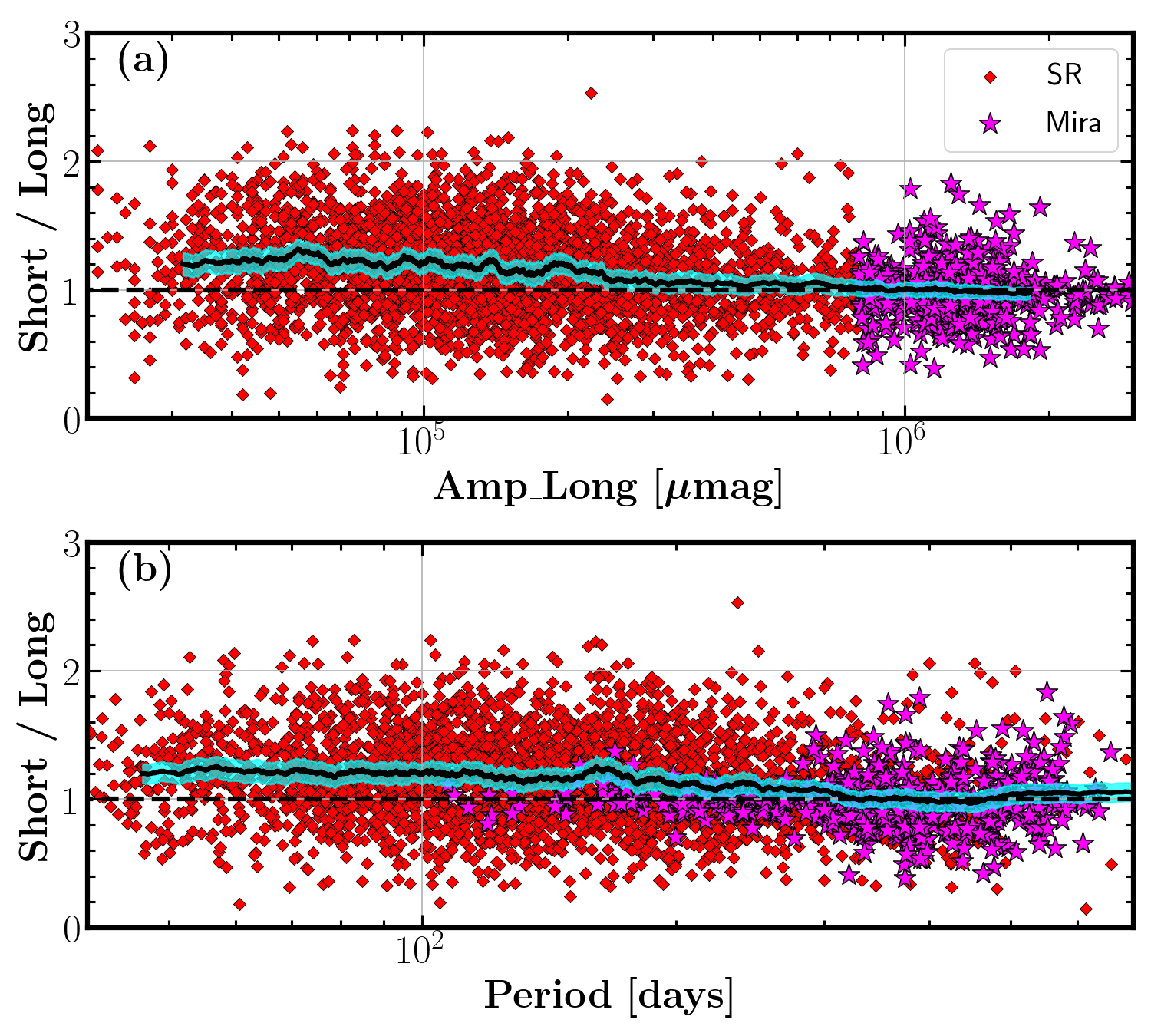}}\\
\caption{\redbf{Similar as Figure \ref{fig:modelifttime} now using the $I$-band light curves of 3383 SRs (red diamonds) and 499 Miras (pink asterisks) in the OGLE-III catalog \citep{soszynski2009a}. These stars are selected to have light curve coverage longer than 10 years and a duty cycle greater than 0.4.}}
\label{fig:modelifttime_ogle}
\end{center}
\end{figure*}

\begin{figure*}
\begin{center}
\resizebox{\textwidth}{!}{\includegraphics{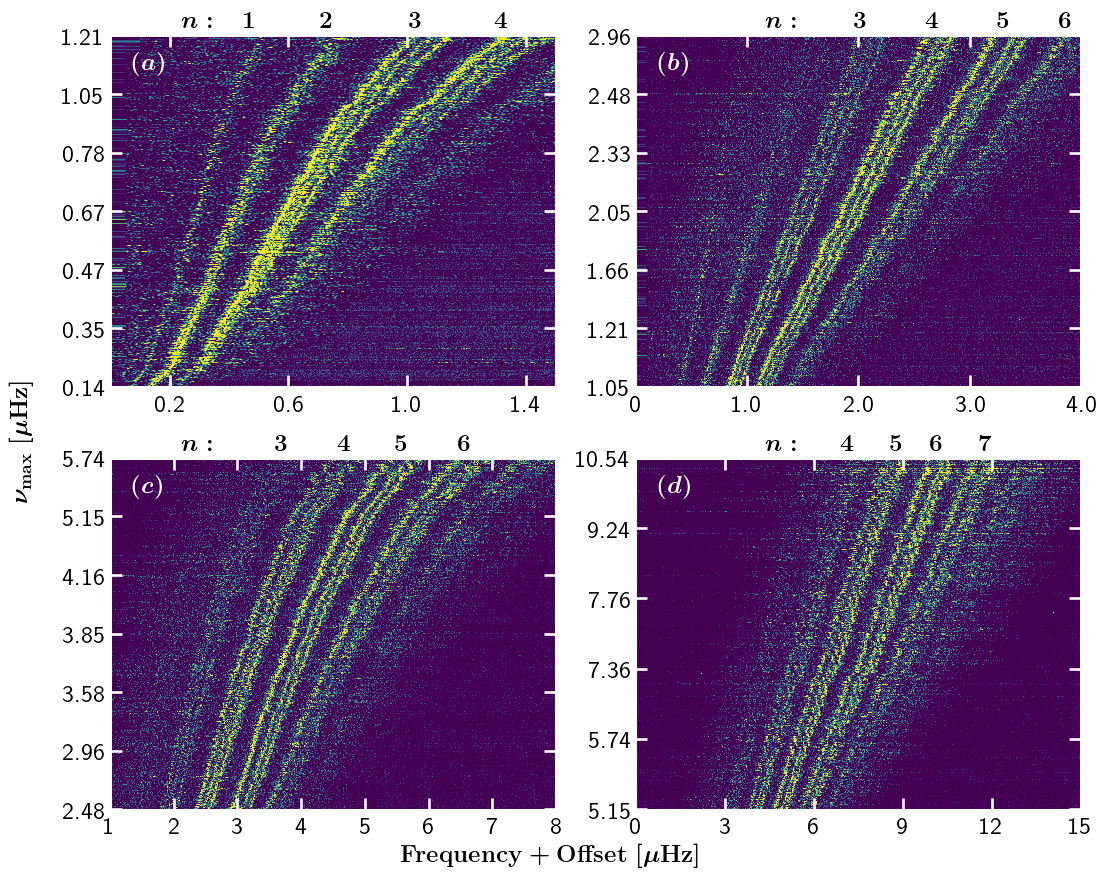}}\\
\caption{Stacked power spectra of high-luminosity red giants with $0.14\ \muHz\leq\numax\leq10.54\ \muHz$. The stacked spectra are shown in four panels so as to highlight in different \numax\ ranges clear ridges associated with multiple angular degrees over several radial orders. Each horizontal band represents one power spectrum with the power color-coded. The ordinate axis is not linear in \numax, hence the different ridge curvatures in the different panels. For each radial order $n\geq3$, as indicated at the top of each panel, $l=1,2,0$ modes lie along the left, middle, and right ridge, respectively.} 
\label{fig:stackedspectra}
\end{center}
\end{figure*}

Figure~\ref{fig:periodamplitude}b illustrates the period-amplitude relation color-coded by the radial order of the dominant mode, which was determined as discussed in Section \ref{OrderAssignment}. We provide radial order assignment only down to $n=2$ using model frequencies from \citet{stello2014a}. 
 We can see that at the boundaries the adjacent radial orders overlap. This
is caused by the stochastic excitation nature of the oscillation modes. For some stars their dominant modes can be non-radial modes and can be a few radial orders away from \numax, which leads to $n$ spanning a larger period range.

We fitted a piecewise linear model and a linear model to the measured relation between the period and amplitude proxy of the dominant mode, and found the former model is preferred due to its smaller Akaike information criterion (AIC) and Bayesian information criterion (BIC). Figure~\ref{fig:periodamplitude} reveals a break point at period $P=4.5$ days, determined by the piecewise linear model fit. Pulsation amplitude increases more rapidly for $P>4.5$ days. 

% \redbf{Note that \citet{banyai2013a} found a break point at $P \simeq 10$ days, and interpreted it as an indication of a transition of the driving mechanisms. Our results do not reveal such a break point. We however find a radial-order transition from $n=4$ to $n=3$ near 10 days.}

Figure~\ref{fig:zerocrossing} shows the dominant mode amplitude as a function of period and as a function of \pextreme\ for \pextreme$>$6 days, respectively. We can see that the parameter \pextreme\ has a tight relation with the amplitude, and is a very good proxy of the dominant mode period (note the similarity of the horizontal axes). All the Miras, two Long Secondary Period (LSP) variables, and three representative SRs are highlighted. We can see that the SRs and LSPs are significantly shifted toward the left in the lower panel while the Miras are much less shifted. This is because SRs and LSPs show more variations in addition to their main periodicity, and thus have more turning points. Figure~\ref{fig:zerocrossing}a shows a significant amplitude decrease at $P\simeq40$ days (the black arrow), which coincides with the radial-order transition of dominant modes from $n=3$ to $n=2$, i.e. from pulsation sequence A to B (see Figures \ref{fig:periodamplitude}b and \ref{fig:kperiod}b). Figure~\ref{fig:zerocrossing}b reveals a sharper lower boundary along the global trend than Figure~\ref{fig:zerocrossing}a. This feature is mainly due to dilution of the oscillation power excess caused by contamination of nearby or foreground/background stars, leading to lower amplitude and smaller \pextreme.

Note that \pextreme\ can be computed from light curves and are hardly affected by \kep\ quarter jumps (by sigma clipping). It could be a robust measure for searching for SRs and Miras observed by the $TESS$ mission \citep{ricker2015a}, in particular those LPVs in the continuous viewing zones. We emphasize that this large and homogeneous sample of LPVs is excellent to study mass-loss triggered by pulsation directly (Yu et al. in preparation). A period of $~60$ days has been argued as a threshold above which substantial dust mass loss is expected \citep[see][and references therein]{mcdonald2018b}. This investigation will be carried out in the second paper of this series.

\section{Period--luminosity relation}
Over the past two decades, one of the major advances in the investigation of LPVs has been the detection of pulsation sequences on the P--L diagram using MACHO and OGLE data. Here, we combine the \kep\ and OGLE LPVs as shown in Figure~\ref{fig:kperiod}. Figure~\ref{fig:kperiod}a shows a Period--$M_K$ diagram of the LPVs in the LMC, where only the dominant period from the OGLE-III catalogue is used \citep{soszynski2009a}. The absolute 2MASS K magnitude, $M_K$, was computed from the Gaia DR2 parallaxes using the same method as \citet{huber2017a} and \citet{berger2018a}. We adopted a LMC distance modulus of 18.54 mag and an extinction $A_V=0.38$ mag \citep{imara2007a}. Sequences A$^{'}$, A, B, C$^{'}$, and C are labeled, following the nomenclature of \citet{soszynski2007a}.  Figure~\ref{fig:kperiod}b shows that the vast majority of the \kep\ $n=3,~4$ LPVs (green and blue) occupy the same region in the  P--L diagram as the sequences A, A$^{'}$. However, the \kep\ results do not show the well-defined sequences of OGLE LPVs. This is more likely due to the approximately six times larger $M_K$ uncertainties, as revealed by Figures~\ref{fig:kperiod}c and \ref{fig:kperiod}d. We do nevertheless expect the pulsation sequences in \kep\ LPVs to be the underlying pattern in  Figure~\ref{fig:kperiod}, given that we have detected radial and non-radial sequences over several radial orders (plotted in a difference way, see Section \ref{orders}).

We calculated the distances in the Period--$M_K$ diagram of all $n=3$ LPVs (most are RGB stars) to sequences b2 and b3, and found they are globally closer to sequence b3 rather than b2. This tells us that sequence A of the OGLE stars possibly corresponds to the second overtone, while the sequences B and C$^{'}$ could be the first overtone. (Note that sequence A comprises sequences $a3$ for AGB stars and $b3$ for RGB stars.) \kep\ Miras are near sequence C. Our findings are consistent with the theoretical results by \citet{trabucchi2017a}, and provide an intermediate solution of radial order assignment over the two sets of contradictory suggestions (see the second open question reviewed in Section~1).  

\begin{figure*}
\begin{center}
\resizebox{1.03\columnwidth}{!}{\includegraphics{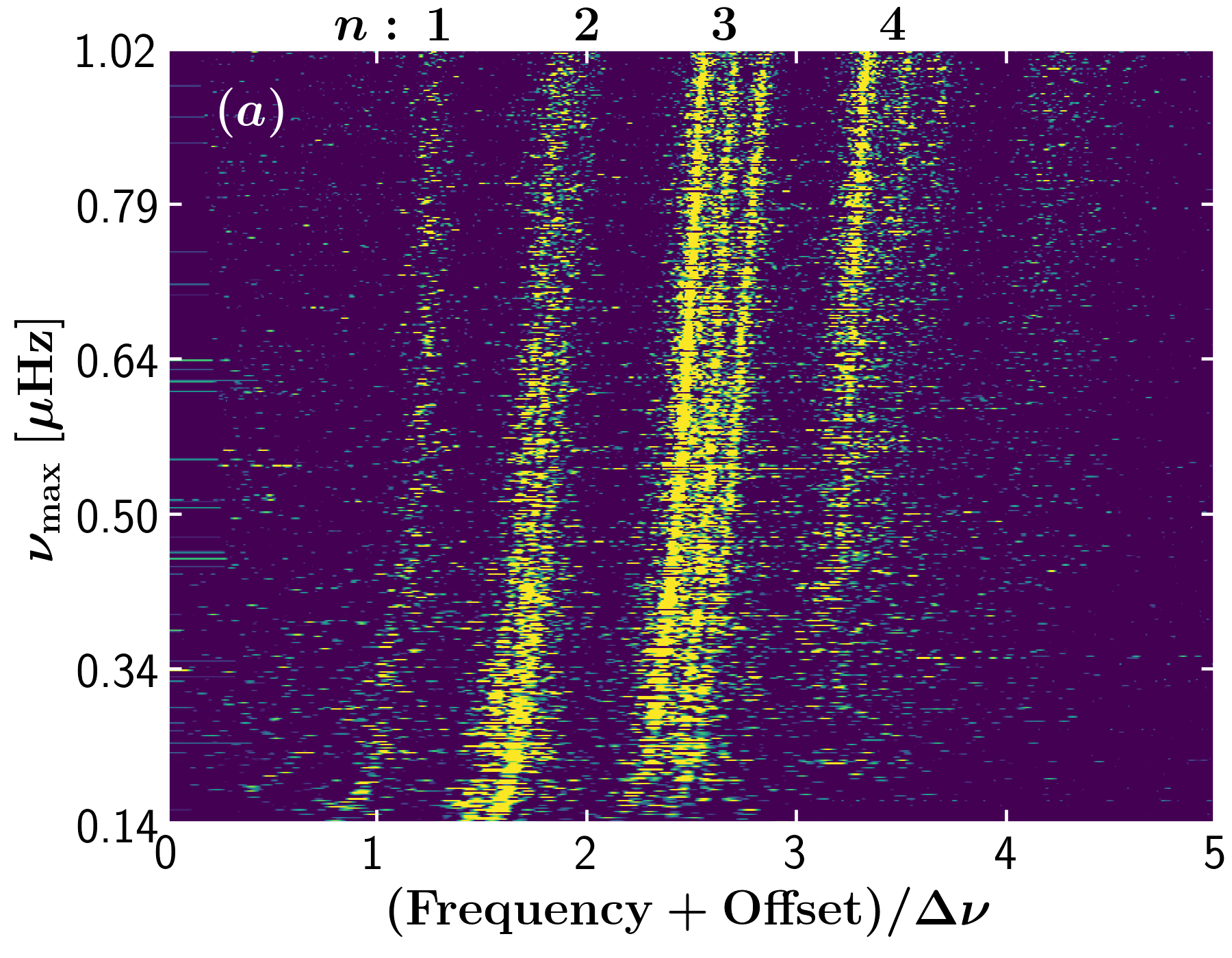}}
\resizebox{1.03\columnwidth}{!}{\includegraphics{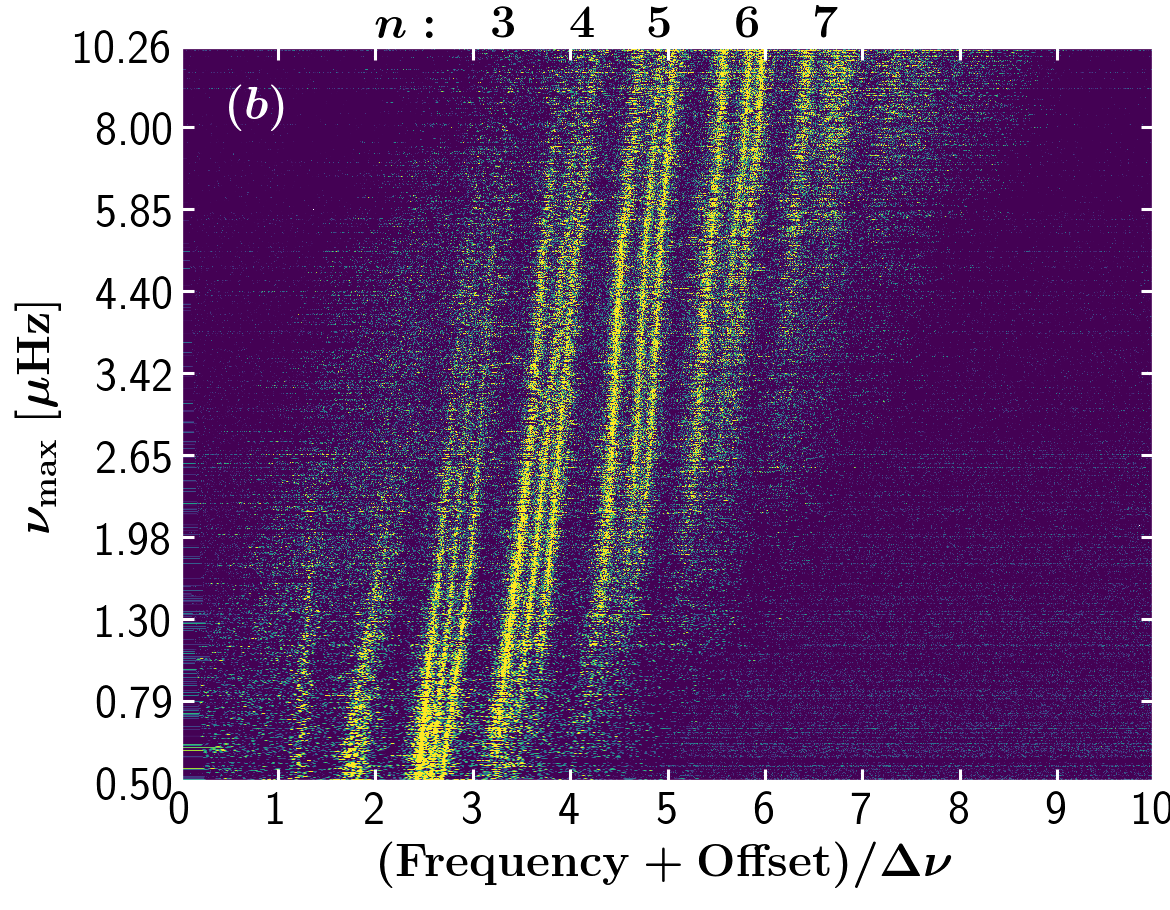}}
\resizebox{1.03\columnwidth}{!}{\includegraphics{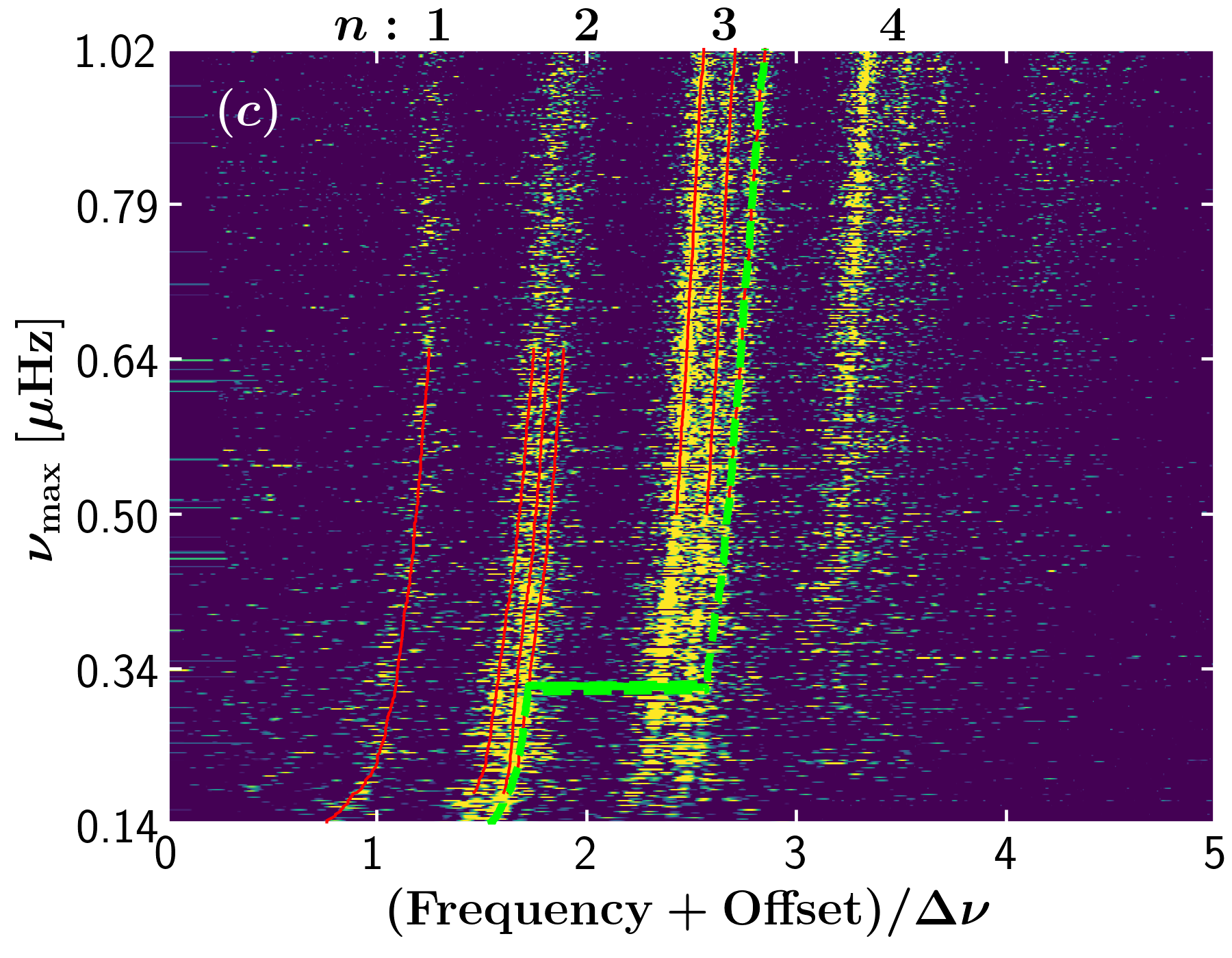}}
\resizebox{1.03\columnwidth}{!}{\includegraphics{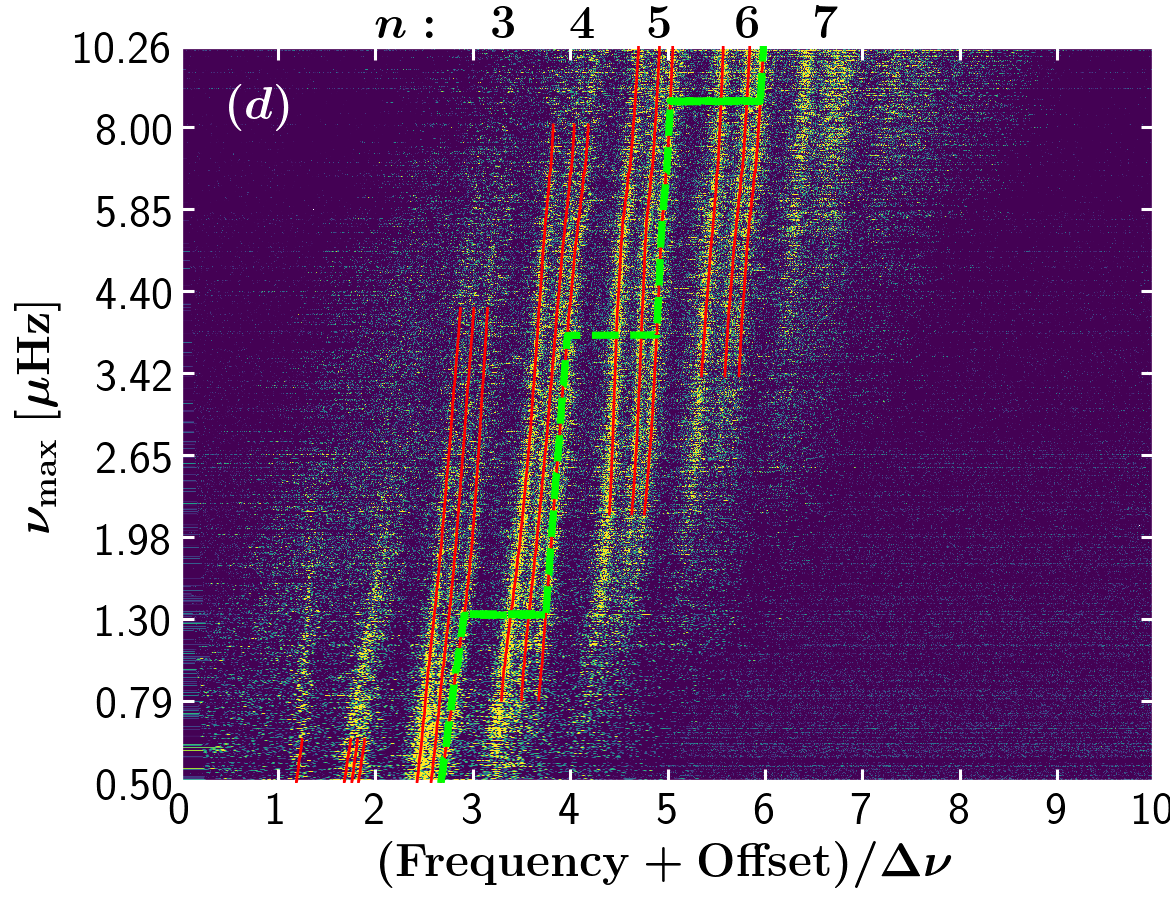}}\\
\caption{Similar to Figure \ref{fig:stackedspectra}, except that the horizontal axis has been divided by \Dnu, and the stacked spectra have been sorted by \numax/\Dnu. Panels \textbf{(a)} and \textbf{(c)} highlight lower radial-order ridges, while Panels \textbf{(b)} and \textbf{(d)} show higher \numax\ values in a wider range. Red lines in the bottom panels indicate $l=1,2,0$ pulsation ridges predicted by model frequencies, which shows an excellent match with the data. Integers on the top of each panel mark radial orders given by model frequencies, which are confirmed by peak-bagging (see Section \ref{peakbagging}). The green dashed lines connect for each star a radial mode that is closest to \numax\ (also shown in Figure~\ref{fig:periodamplitude}b).}
\label{fig:ridges}
\end{center}
\end{figure*}

\section{Stochastic vs Mira-like excitation in Semiregular variables}
\label{excitation}
To address the question of mode excitation in SRs, we consider the properties characterizing solar-like oscillations. If a detected mode is resolved into a Lorentzian profile, its amplitude in the Fourier spectrum as normalised here (see the caption of Figure~\ref{fig:lcs}) decreases with increasing length of the time series. However, if a detected mode is unresolved, and hence can be described by a sinc function, its amplitude does not depend on the length of the time series. Bearing this in mind, we cut the total light curve for each star into three segments with equal length, and selected the one with the highest duty cycle to measure the pulsation period and amplitude in the same way as before (see Section~\ref{lpvperiodamp}).

Figure \ref{fig:modelifttime} shows a comparison of the dominant mode amplitudes, measured from the full-length and 1/3-length light curves. We observe clearly a systematic offset in the amplitude ratios when the amplitude is less than 0.1 mag, or $P\simeq70$ days, and good consistency in amplitude for the Miras. The scatter in the amplitude ratios is significantly larger for the SRs than for the Miras. The offset confirm that the pulsations are stochastically excited in SRs and self-excited in Miras \citep{christensen-dalsgaard2001a, bedding2003a}. 

The expected amplitude ratio is $\sqrt{3}$  if the modes are completely resolved, and unity if the modes are unresolved. Our measured ratio for the SRs is $\sim$1.2, which is smaller than the expected value, indicating the modes are partially resolved. The ratio decreases slowly with increasing amplitude, showing that the intrinsic mode lifetimes increases slowly.

To test whether the systematic offset of the amplitude ratio is statistically significantly, we performed a two-sided one sample t-test against the stars with the amplitude smaller than 0.1 mag. Our null hypothesis was that there is no significant difference between the ``short" and ``long" amplitudes, namely the amplitude ratio is unity. We found that the t-statistic value is 26.7, the degrees of freedom is 3139, and the p-value is $2.1 \times 10^{-141}$. The p-value is far less than the significance level $\alpha=0.05$, so we rejected the null hypothesis. This means that the ``short" amplitudes are significantly larger than the ``long"  amplitudes in a statistical sense.

\redbf{We note that for the \kep\ Miras and longest period SRs, the \kep\ light curves cover only a few cycles. Thus, we investigated the amplitude ratio of SRs and Miras in the OGLE III catalog that have light curve coverage longer than 10 years and a duty cycle greater than 0.4. Using the same method as for the \kep\ LPVs, we found very similar offsets of the amplitude ratio, as shown in Figure \ref{fig:modelifttime_ogle}, and therefore further supports the conclusion that we see a transition from predominantly stochastic-excitation in SRs to a more coherent  self-excition mechanism in Miras as we found from the \kep\ LPVs.}

\begin{figure*}
\begin{center}
\resizebox{0.9\textwidth}{!}{\includegraphics{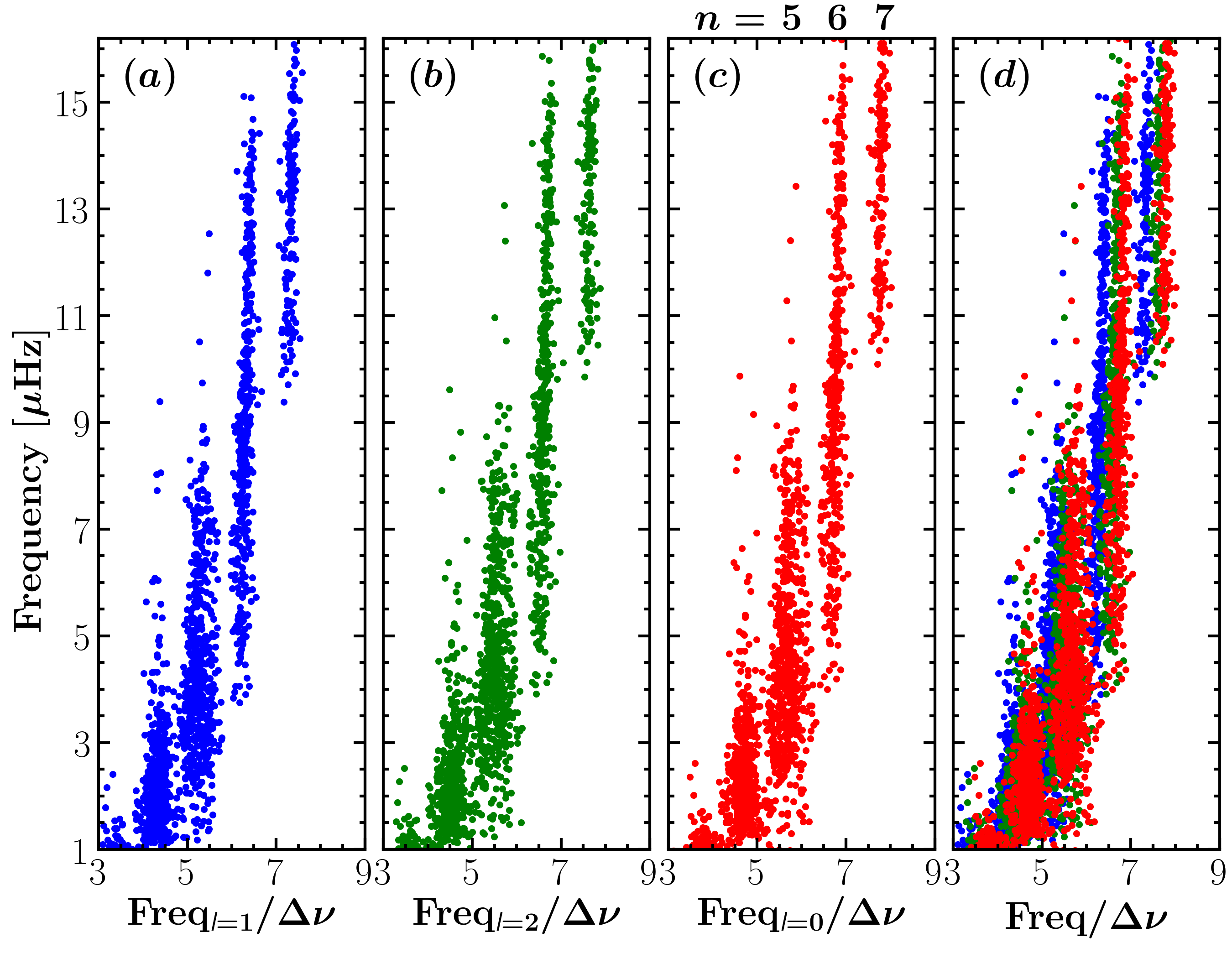}}\\
\caption{Oscillation patterns for (a) $l=1$, (b) $l=2$, and (c) $l=0$ modes of high-luminosity red giants. (d) The combination of the $l=1,2,0$ oscillation patterns. The horizontal axes are the frequency divided by measured \Dnu, while the vertical axes are frequency. For each star only one $l=1$, 2, and 0 mode are shown.}
\label{fig:pattern}
\end{center}
\end{figure*}

\begin{figure*}
\begin{center}
\resizebox{0.9\textwidth}{!}{\includegraphics{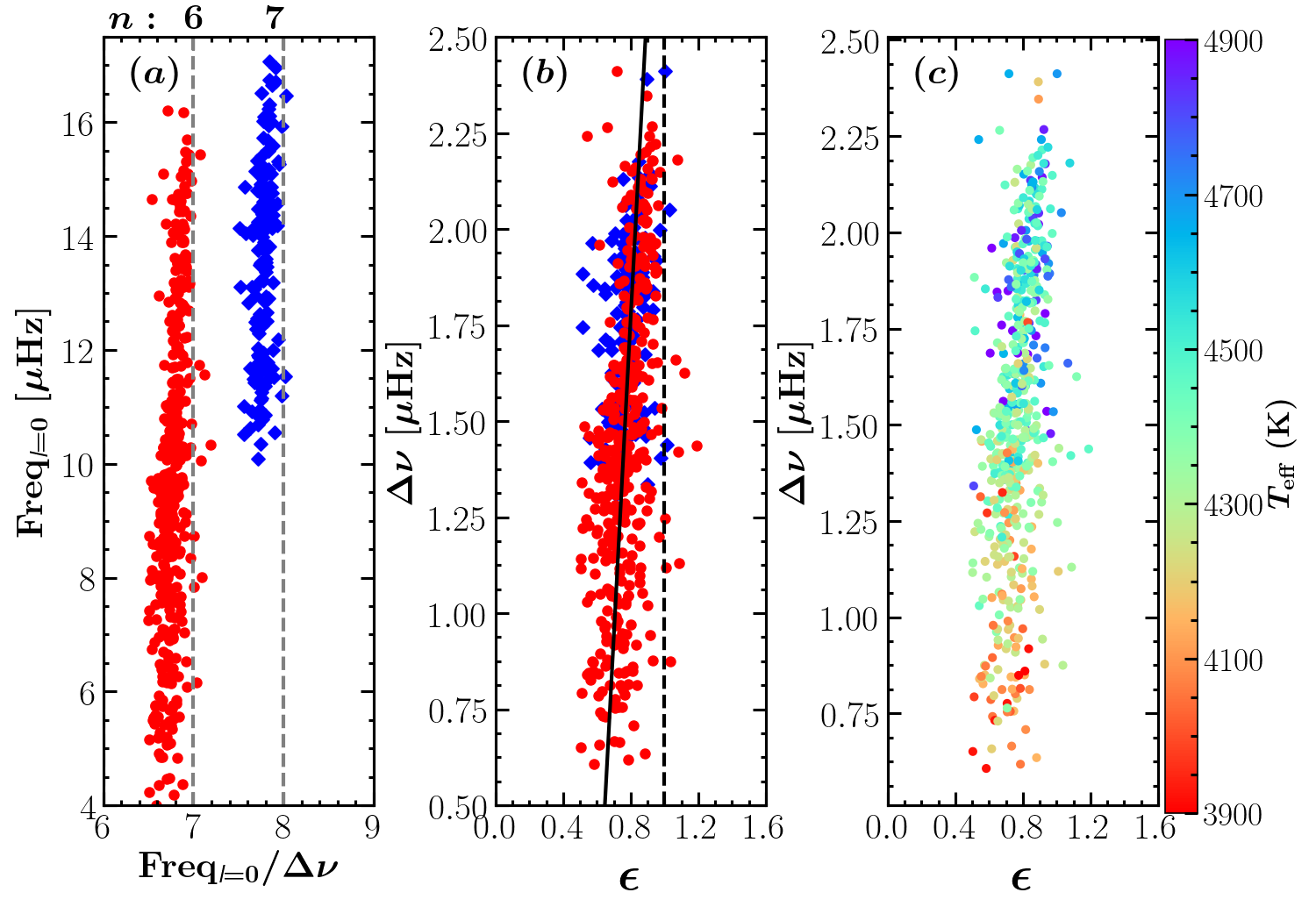}}\\
\caption{(a) Same as Figure~\ref{fig:pattern}c now only for the two ridges for which we aim to measure their radial orders. (b) The relation between \Dnu\ and $\epsilon$, where $\epsilon$ is measured via Equation \ref{asymptotics} by assigning a radial order of 6 to the ridge in red and 7 to the ridge in blue. (c) Same as (b) now color-coded by effective temperature.}
\label{fig:epsilon}
\end{center}
\end{figure*}

\section{radial order assignment}\label{orders}
\subsection{Radial order assignment by theoretical frequencies}
\label{OrderAssignment}
Figures \ref{fig:stackedspectra} and \ref{fig:ridges} show the stacked power spectra of 2000 stars with clearly detected oscillations, for which the \numax\ values are in the range 0.14 \muHz\ $\leq$ \numax$ \leq$ 10.54 \muHz\ (period 1.1-82 d). For this, we used dedicated procedures (see below) to detect the clear pulsation ridges corresponding to multiple angular degrees $l=0,1,2$ for a number of radial orders $1\leq n \leq 7$. The basic idea in constructing these ridges is to construct a series of template power spectra as references, and then shift an observed target power spectrum to align with the references. The method is summarized as follows.

\begin{enumerate}
    \item We used the SYD pipeline \citep{huber2009a} to prepare background-divided power spectra, and to measure global seismic parameters, \numax\ and \Dnu.
    
    \item We then created template spectra using the model frequencies from \citet{stello2014a}. They calculated the stellar models for a fixed stellar mass of 1$\rm M_{\odot}$ \footnote{This approaches a median mass of 1.2 $\rm M_{\odot}$ for our sample.} at solar metallicity, using the MESA \texttt{1M\_pre\_ms\_to\_wd} test suite case \citep{paxton2011a, paxton2013a} \footnote{A detailed description of the implemented physics is given by \citet{stello2014a} and references therein.}. Their adiabatic frequencies were calculated using ADIPLS \citep{christensen-dalsgaard2008a}. Both \numax\ and \Dnu\ were derived from seismic scaling relations for each stellar model. From the frequencies of a given stellar model, we built a template spectrum, where each mode was described by a Lorentzian profile and its height was modulated by a Gaussian envelope. The Gaussian envelope was centered at \numax\ with a standard deviation of \Dnu, hence FWHM $\simeq$ 2.4 \Dnu. Both \numax\ and \Dnu\ were subsequently interpolated with 100-times finer step sizes, as well as the associated model frequencies. The model frequencies from \citet{stello2014a} were only available for models with $\numax~\geq~0.2~\muHz$. We linearly extrapolated the model frequencies for \numax\ down to 0.1 \muHz. This extrapolation only involved 20 (0.8\%) stars, and the clear ridges for these stars shown in Figures \ref{fig:stackedspectra}a and \ref{fig:ridges}a ensure its reliability to reveal the regular frequency patterns. 
    
    \item For each target spectrum, we searched for its best-matching template spectrum by choosing the one whose maximum \mbox{cross-correlation} with the target spectrum is the largest. We then shifted the target spectrum with respect to the best-matching template spectrum by an offset equal to the lag of the maximum cross-correlation. When the shifted spectra were stacked they form clear ridges, as seen in Figure \ref{fig:stackedspectra} and \ref{fig:ridges}. Note that a spectrum was not used if the shift was greater than 1/2 \Dnu, to avoid the observed and template spectra being mismatched by one or more radial orders (only 7 stars were discarded in this step). From the best-matching template spectrum, its theoretical frequencies were used to determine the radial order and angular degree for the dominant mode, which were used in Section \ref{lpvperiodamp}.
    
    \item Lastly, we sorted the shifted observed spectra by \numax\ for Figure \ref{fig:stackedspectra} and by \numax/\Dnu\ for Figure \ref{fig:ridges}. Here, the values of \numax\ and \Dnu\ the ones corresponding to the best-matching template spectra. 
\end{enumerate}

\begin{figure*}
\begin{center}
\resizebox{\textwidth}{!}{\includegraphics{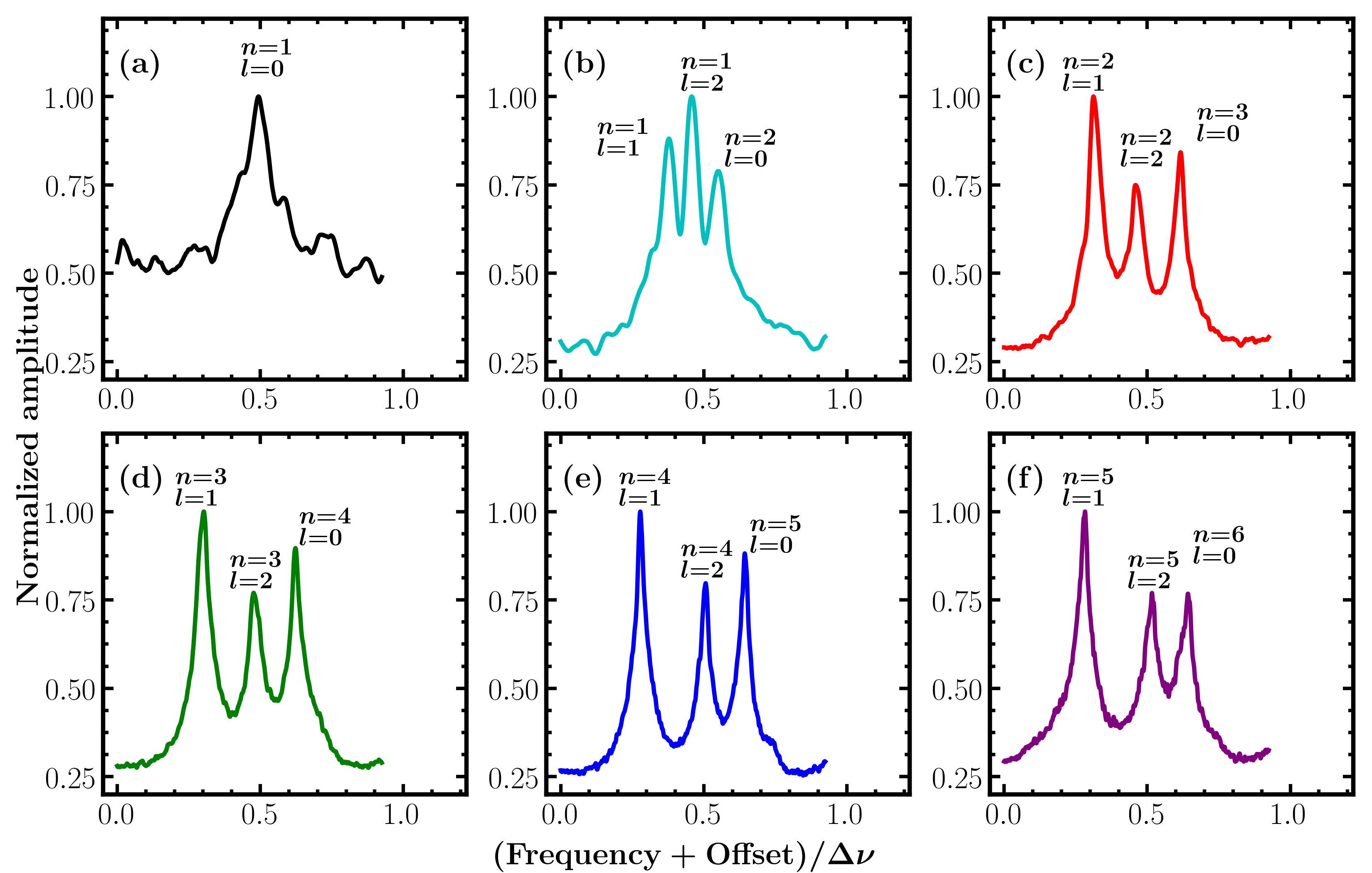}}\\
\caption{Relative amplitude of $l=1,2,0$ modes across various radial orders, as indicated by the legends. As an example, the three peaks from left to right in Panel (\textbf{b}) correspond to $(n, l)$=(1,1), (1,2), and (2,0) modes, respectively. Dipole modes are dominant in the radial orders of $n=3,4,5,6$, while quadrupole modes are dominant in the radial order of $n=2$. The triplet structure shrinks gradually toward lower radial orders, and merges eventually in the radial orders of $n=1$.}
\label{fig:amplitude}
\end{center}
\end{figure*}

Figure \ref{fig:stackedspectra} displays the radial orders of $l=0$ modes measured from the model frequencies as indicated at the top of each panel (the values of $n$ hereafter refer to the radial order of $l=0$ modes, unless specifically stated).  We have used an independent method to confirm this radial order assignment, which will be presented in the next section. \citet{stello2014a} firstly recognized triplet structures that consist of dipole ($l=1$) modes to the left, quadrupole ($l=2$) modes in the middle, and radial ($l=0$) modes to the right. The so-called $f$-mode ridge \citep{stello2014a}, which would lie to the left of the ridge $[n,l]=[1,0]$, is not clearly detected. In Figure \ref{fig:stackedspectra} we can see the triplet structure is gradually resolved towards larger \numax\ values and higher radial orders. 

We again show the stacked power spectrum in Figure~\ref{fig:ridges}, now with the horizontal axes divided by \Dnu. Figure~\ref{fig:ridges} shows sub-ridges related to multiple angular degrees. Here, we can see the ridge of the first radial overtone ($n=2$) is marginally resolved in the range $0.45\ \muHz\leq\numax\leq0.64\ \muHz$, and the individual components of the triplet structure gradually merge towards the lower-\numax\ end. Longer light curves are thus required to resolve triplet structures for the radial order $n=2$ and $\numax<0.45\muHz$, and for the entire ridge of the radial fundamental mode. The red lines in the lower panels of Figure~\ref{fig:ridges} indicate the model frequencies, which match the observations well. 

The well-resolved ridges separated by the so-called small and large frequency separations for various radial orders at such late evolutionary phase (\numax\ down to 0.14 \muHz) resemble the ridges seen in less-luminous red giants \citep[e.g.][]{mosser2011a}. This implies that the asymptotic relation of acoustic modes remains helpful for assigning radial orders and angular degrees, although the relation is expected to break down at low radial orders. The asymptotic relation is given as \citep{tassoul1980a}: 
\begin{equation}
\label{asymptotics}
\nu_{n,l}=\Delta\nu(n+\frac{l}{2}+\epsilon)-\delta\nu_{0,l},
\end{equation}
where $\nu_{n,l}$ is the eigenfrequency at radial order $n$ and angular degree $l$, $\epsilon$\ is an offset parameter, and $\delta\nu_{0,l}$ is the small frequency separation between radial and non-radial modes.

\subsection{Radial order assignment by peak-bagging}
\label{peakbagging}
In order to confirm the radial order assignment given by model frequencies, we aim to determine the radial orders from the observations by measuring mode frequencies, \Dnu\ and $\epsilon$. For this, one of the most important steps is to make good initial guesses of mode frequencies. Here, for each star we used the frequencies from the best-matching template spectrum, as defined in Section \ref{OrderAssignment}. We then fitted three Lorentzian profiles to the $l=1, 2, 0$ modes. Only the triplet structure with the largest power was picked.

Figure~\ref{fig:pattern} shows oscillation patterns of $l=1, 2, 0$ modes for stars with \numax\ > 1.0 \muHz. This is analogous to the pattern seen in higher-\numax\ red giants \citep{bedding2010a, huber2010a, mosser2011a}, here restricted to low-\numax\ stars. The horizontal scatter increases with decreasing radial order, which is mainly because the mean large frequency separation \Dnu\ is less well defined towards lower-\numax\ stars. For these low-\numax\ stars, only a few orders of modes are excited, and modes with the same angular degrees deviate significantly from equally spaced).

To assign a radial order using Equation~\ref{asymptotics}, we started with the two ridges on the far right in Figure~\ref{fig:pattern}c ($n=6,7$). This is because (1) for radial modes, the small frequency separation term, $\delta\nu_{0,l}$, is zero; (2) the associated \Dnu\ can be measured more precisely, as indicated by the much smaller scatter, compared to the $n=$ 4 and 5 ridges; (3) the asymptotic relation works more accurately at higher radial orders; and (4) the lower radial order can be easily deduced once higher radial orders are identified.

Figure~\ref{fig:epsilon}a shows the two radial-mode ridges with \numax\ in the range $4\ \muHz<\numax<17\ \muHz$. Figure~\ref{fig:epsilon}b shows \Dnu\ as a function of the offset $\epsilon$ that was computed by using Equation~\ref{asymptotics} and by assuming $n=6$ for the red ridge and $n=7$ for the blue ridge. Clearly, the $\epsilon$ values are collectively smaller than unity, which is in agreement with \citet{mosser2011a} and \citet{kallinger2012a} for the stars in their samples that overlap in \numax\ with ours. This result confirms the ridges in red and blue correspond to the radial orders of 6 and 7, respectively. This radial order assignment is thus consistent with the assignment given by the model frequencies as shown above. Figure~\ref{fig:epsilon}c shows that for our sample the offset $\epsilon$ is an increasing function of \Dnu, and \Dnu\ is an increasing function of effective temperature (see the colourbar). This means $\epsilon$ increases with increasing effective temperature. This relation between the offset $\epsilon$ and effective temperature has also been found in both dwarfs and giants \citep{white2011a, white2012a, lund2017a}. With $n=6$ and $n=7$
confirmed, the identifications for $n=1$ to 5 follows from the discussions in Section \ref{OrderAssignment}.

To summarize, from the analyses in Section \ref{orders} and the comparison between the \kep\ and OGLE LPVs (Figure~\ref{fig:kperiod}b), we provide a solution to the open question on the radial order assignment of LPVs as detailed in the introduction. Our results confirm that the radial orders of $n$=1, 2, 3, and 4 can be used to explain the sequences C, C$^{'}$ and B combined, A, and A$^{'}$ in the P--L diagram, respectively, which is consistent with the recent theoretical explanations by \citet{trabucchi2017a}.

\section{Are dipole modes dominant among the pulsations of LPVs?}\label{dominantmodes}
Which modes are dominant, radial or non-radial modes? To answer this question, we measured relative amplitudes of $l=0,1,2$ modes for the radial orders $1\leq n \leq 6$. 

In order to measure the amplitude for a given $n$ and $l$, we used the stacked power spectrum, as shown in Figure~\ref{fig:ridges}, and summed up the amplitude along the associated ridge indicated by the red lines in Figures~\ref{fig:ridges}c and \ref{fig:ridges}d. The collapsed total amplitude was evaluated over the \numax\ ranges equal to the length of the red lines. For each star, its background-divided power spectrum was normalized so that the amplitude of the highest peak was unity. Lastly, for each radial order, the collapsed amplitude was normalized to set its highest peak to unity. The results are shown in Figure \ref{fig:amplitude}. 

Figure \ref{fig:amplitude}a shows only a single peak for $n=1$. Interestingly, for $n=2$ (note radial orders are defined for $l=0$ modes in this work), the middle peak ($l=2$ modes) is globally the highest. This property is distinct from the higher radial orders $3\leq n \leq 6$, for which all the collectively dominant modes are dipole modes . Note that, as shown in Figure~\ref{fig:ridges}c, for radial order $n=2$ the dominant $l=2$ mode is marginally visible at \numax~$\simeq$ 0.55 \muHz, and the $l=2,0$ ridges are gradually merged as \numax\ decreases.

\begin{figure*}
\begin{center}
\resizebox{0.95\textwidth}{!}{\includegraphics{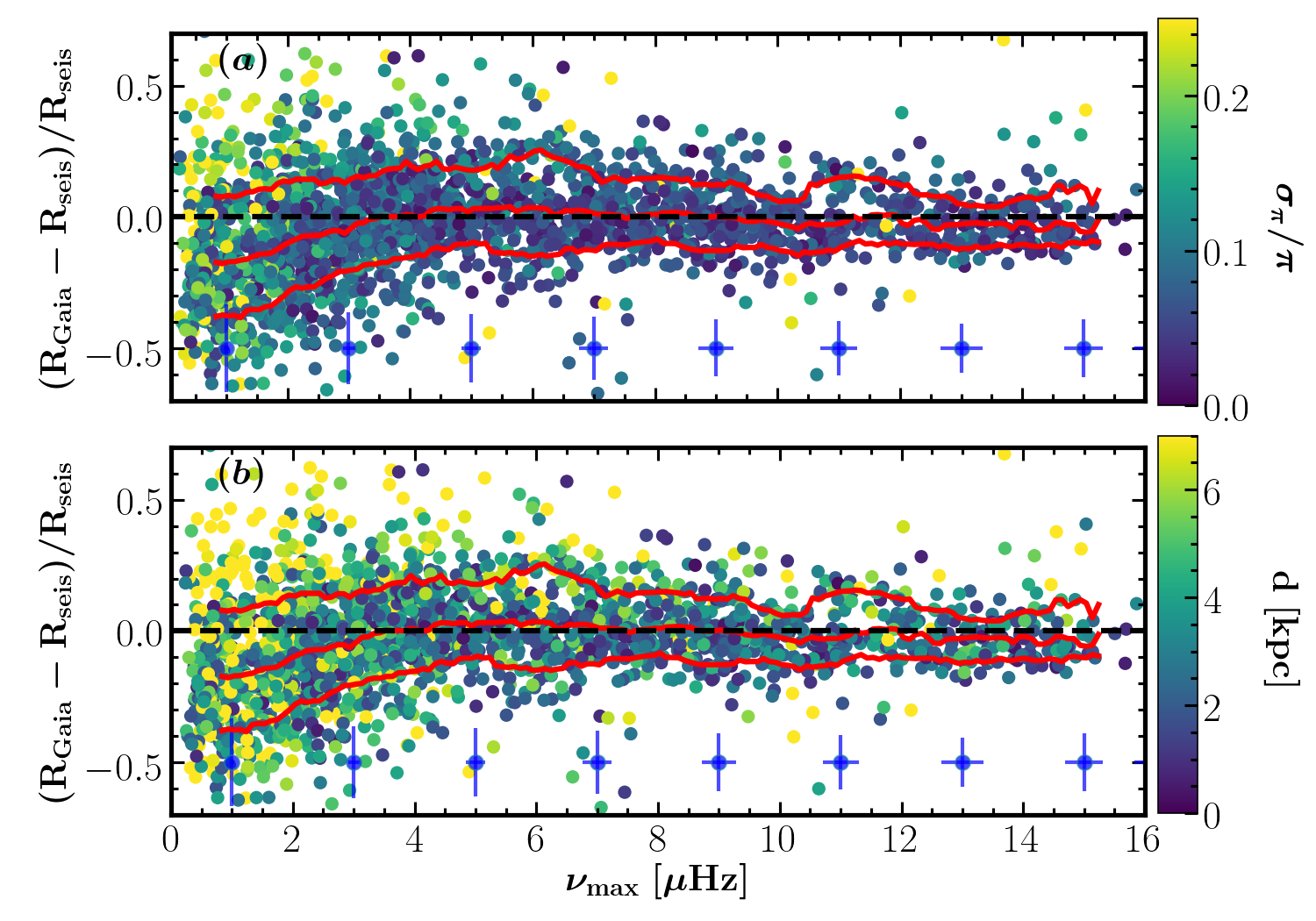}}
\caption{(a) Relative radius difference as a function of \numax, color-coded by (a) the fractional uncertainty of Gaia DR2 parallax and (b) distance derived using the method by \citet{huber2017a} and  \citet{berger2018a}. The 16th, 50th, and 84th percentiles are indicated by the red lines in each panel. The median uncertainty in each \numax\ bin in steps of 2 \muHz\ are indicated.}
\label{fig:radiicompnumax}
\end{center}
\end{figure*}

Our result yields an average $l=1$ visibility $V^2_1=$1.5 (averaged over n=2, 3, 4, 5, and 6) and an average $l=2$ visibility $V^2_2=$ 0.7 (averaged over n = 3, 4, 5, and 6), and thus are  consistent with the theoretical prediction ($V^2_1=$1.5, $V^2_1=$0.6) and with the observed values ($V^2_1=$1.3, $V^2_1=$0.6) for the solar-like oscillations in red giants \citep[see][and references therein]{mosser2012a}.

Our findings of the dominant $l=1$ modes with radial order $n \geq 3$ are consistent with the results by \citet{mosser2013a}(see their Figure 9) and \citet{stello2014a}. However, the findings of the dominant $l=2$ modes in radial order $n=2$ (at least in a higher \numax\ range) is discrepant to \citet{mosser2013a} who argued that radial modes are dominant when \numax~$\lesssim$~1.0~\muHz. We note \citet{stello2014a} clearly detected the $n=2$ ridge but did not resolve the associated sub-ridges.

Another interesting feature shown in Figure \ref{fig:amplitude} is that the triplet structure gets more narrow with decreasing radial order. This feature makes it difficult to resolve multiple angular degrees, given the 4-year baseline of \kep\ light curves. The closely spaced triplet structures are very different in the LPVs than in less-luminous red giants. For the latter, $l=1$ modes are nearly located at the midpoint of adjacent $l$ = 0 modes \citep[e.g.][see their Figure 10]{huber2010a}. OGLE data will be valuable for studying the unresolved or marginally resolved triplet structures, thanks to 4 years of data set from the OGLE-II project and 8 years of observations from the OGLE-III project. This will be presented in a future paper.

\section{Testing seismic scaling relations for high-luminosity red giants}
\label{testseis}
 Since the seismic scaling relations provide an efficient way to derive stellar fundamental properties, such as mass and radius, their validity has been extensively tested on dwarfs and giants \citep[see][for reviews]{chaplin2013b, hekker2017a}. It seems inevitable that the seismic scaling relations should break down at a certain evolutionary stage, at least in Mira variables in which pulsations are self-excited via a heat-engine mechanism, which is different from solar-like oscillations. We will test the seismic scaling relations for high-luminosity red giants.

\begin{figure*}
\begin{center}
\resizebox{\columnwidth}{!}{\includegraphics{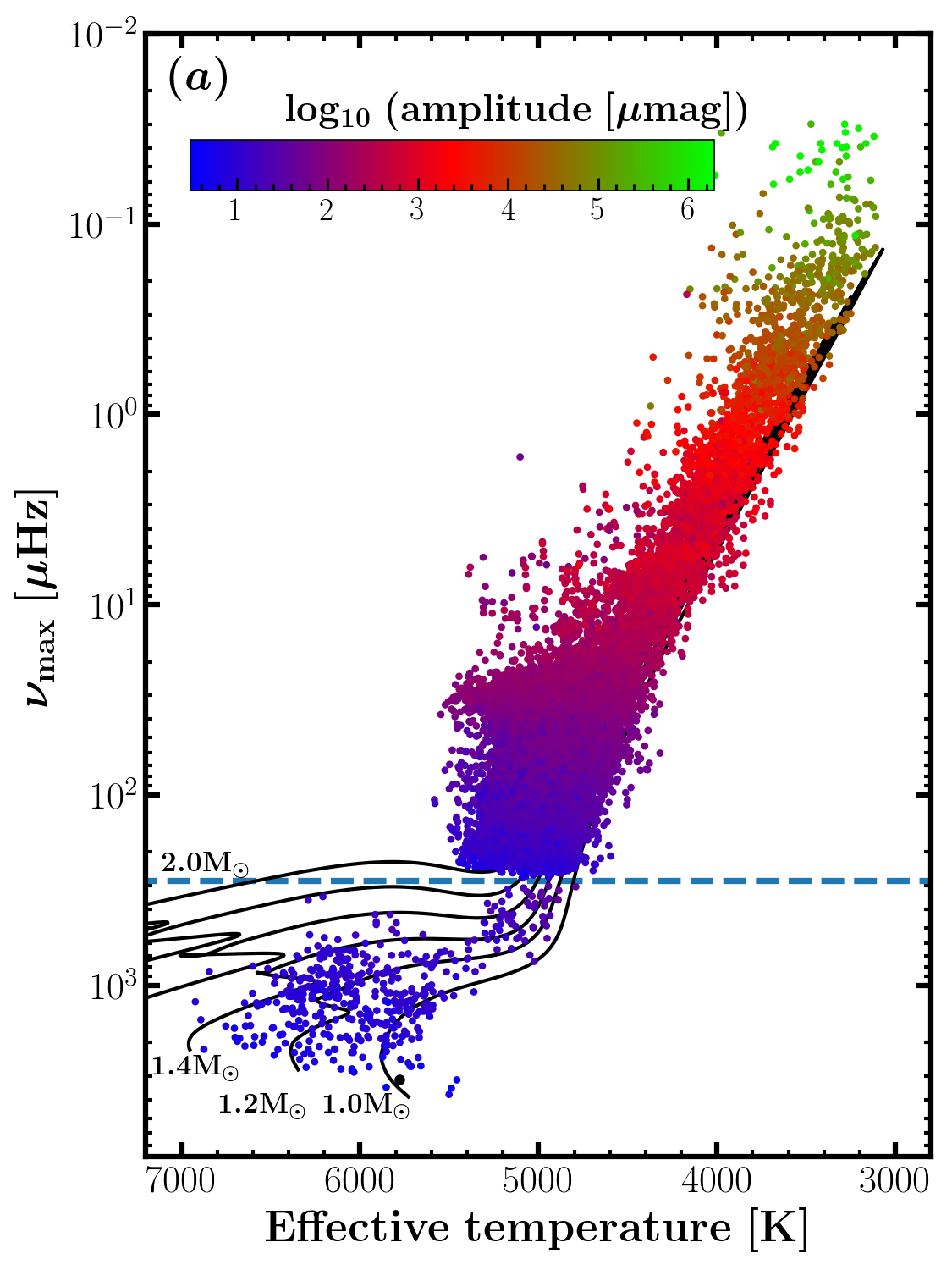}}
\resizebox{\columnwidth}{!}{\includegraphics{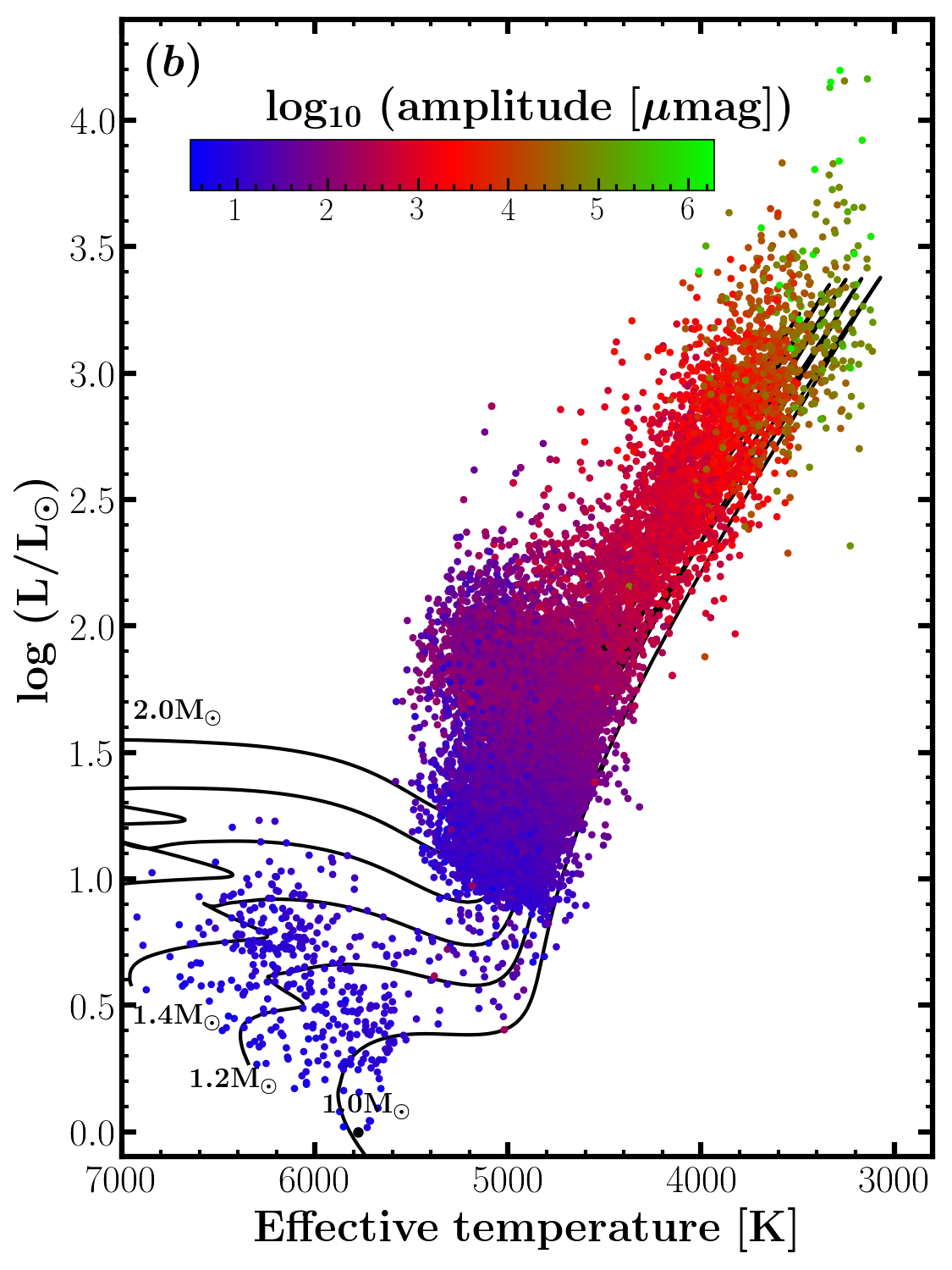}}
\caption{\textbf{(a)} Seismic H-R diagram (\numax\ vs \teff). The estimates of \numax\ and oscillation amplitude were adopted from \citet{huber2011a} for main-sequence and subgiant stars, \citet{yu2018a} for low/intermediate luminosity red giants, and this work for high-luminosity red giants (\numax < 15 \muHz). The color bar indicates the oscillation amplitude of the radial mode for the stars from the literature, and the amplitude proxy for the stars analysed in this work. We take effective temperatures from \citet{mathur2017a}, and update them wherever temperature is cooler than 3200 K (for the scheme see the text). The dashed line indicates the long-cadence Nyquist frequency, below which it is very challenging to detect the oscillations using the \kep\ long cadence data. \textbf{(b)} H-R diagram. The values of luminosity are calculated from Gaia DR2 parallaxes, either from this work or \citet{berger2018a}.}
\label{fig:hr}
\end{center}
\end{figure*}

 Figure~\ref{fig:radiicompnumax} shows a comparison between radii calculated from the seismic scaling relations and radii derived from the Gaia DR2 parallaxes \citep{lindegren2018a}. To calculate the seismic radii, we use the following relation:

 \begin{equation}
     \frac{R}{R_{\odot}} \simeq \left(\frac{\numax}{\numax_{\odot}}\right) \left(\frac{\Dnu}{\Dnu_{\odot}}\right)^{-2} \left(\frac{\teff}{\teff_{\odot}}\right)^{1/2},
\label{seismicrelation}     
 \end{equation}
where the solar references are $\numax_{\odot}=3090~\muHz$, $\Dnu_{\odot}=135.1~\muHz$ \citep{huber2011a}, and $\teff_{\odot}=5777~$K. The global seismic parameters, \numax\ and \Dnu, were measured using the SYD pipeline \citep{huber2009a}, which gave very good agreement with \citet{yu2018a} (The mean differences were 0.8\% in \numax\ and 0.2\% in \Dnu\ for 692 stars in common) and \citet{pinsonneault2018a} (The mean differences were 1.3\% in \numax\ and 0.2\% in \Dnu\ for 531 stars in common). Effective temperatures in this work were adopted from \citet{mathur2017a}, which were mainly based on photometry, and are also consistent with APOGEE spectroscopic temperatures from \citet{pinsonneault2018a}. We then derived radii from Gaia DR2 parallaxes using the same method as \citet{huber2017a} and \citet{berger2018a}. We applied a cut to fractional parallax uncertainty, namely, $\sigma_{\pi}/\pi$ < 0.6. From this, we obtained a sample of 2241 LPVs with both seismic and \mbox{parallax-based} radii available.

Figure~\ref{fig:radiicompnumax} shows the relative radius difference as a function of \numax, colour-coded by the fractional uncertainty of Gaia DR2 parallax, $\sigma_{\pi}/\pi$, and distance. We can see that the radii derived from the two independent methods are consistent when \numax~$\gtrsim$~3~\muHz. This suggests that the seismic scaling relations remain accurate for luminous red giants (\numax~$\gtrsim$~3~\muHz, R~$\lesssim$~40 R$_{\odot}$, or \mbox{log $\rm L/L_{\odot}$~$\lesssim$~2.6}). \citet{lindegren2018a} suggested Gaia DR2 parallaxes are too small by 0.03 mas, thus we added this systematic offset to the parallaxes for individual stars when computing their parallax-based radii. This remarkable consistency confirms the existence of the 0.03 mas zero-point offset in Gaia DR2 parallaxes. We also tested the parallax zero-point correction method proposed by \citet{zinn2019a}, where the median correction value was 0.065 mas for our sample. The results shows that the parallax-based radii were 7\% smaller than the seismically derived radii. The discrepancy in the different parallax zero-point corrections might be linked to the different star samples where our stars are more luminous than those used by \citet{zinn2019a} for their calibration.

Figure~\ref{fig:radiicompnumax} also shows that the relative radius difference increases with decreasing \numax\ when \numax~$\lesssim$ 3 \muHz. This increasing systematic offset suggests that the seismic scaling relations is likely to gradually break down at \numax~$\lesssim$ 3 \muHz. This \numax\ threshold tells us that the seismic scaling relations may not be applicable to SRs, which typically have a \numax\ $<$ 0.5 \muHz. 

From Figure~\ref{fig:radiicompnumax}, we can see that the dispersion is mainly caused by the uncertainties. We carried out two-sided one sample t-tests to confirm the validity of the seismic scaling relations in the regimes $\numax\ > 3\ \muHz$ and $\numax\ \leq 3\ \muHz$.

In both cases, our null hypothesis was that the seismic and Gaia-based radii are not significantly different. In other words, we hypothesized $\rm ( R_{Gaia}-R_{seis})/R_{seis}$ = 0. For the $\numax\ > 3\ \muHz$, we found that the t-statistic value is -0.18, which we used to calculate a p-value along with the degrees of freedom $n-1=1342$. The p-value is 0.85, which is far greater than the significance level $a=0.05$, so we accepted our null hypothesis. This means there is no statistically significant difference between the seismic and Gaia-based radii.

For $\numax\ \leq 3\ \muHz$, we found that the t-statistic value is -13.94. Together with the degrees of freedom $n-1=888$, we obtained a p-value of 4.01$\times 10^{-40}$. The p-value is far less than the significance level $a=0.05$, so we rejected the null hypothesis. This means the seismic radii are statistically significantly different from the Gaia-based radii.

%%%%%%%%%%%%%%%%%%%%%%%%%%%%%%%%%%%%%%%%%%%%%%%%%%%%%%%%%%%%%%%%%%%%%%%%%%%%%%%%%%%%%%%%%%%%%%%%%%
%%%%%%%%%%%%%%%%%%%%%%%%%%%%%%%%%%%%%%%%%%%%%%%%%%%%%%%%%%%%%%%%%%%%%%%%%%%%%%%%%%%%%%%%%%%%%%%%%%
\section{Hertzsprung-Russell Diagram of \kep\ oscillators}
Figure~\ref{fig:hr}a shows a seismic H-R diagram, color-coded by the oscillation amplitude per radial mode. We note that for the stars in our sample with \teff\ < 3200K from \citet{mathur2017a}, \teff\ was poorly determined. For this, we updated their temperatures by using g-K$_s$ colour calculated from SDSS g and 2MASS K$_s$ magnitude and following the empirically calibrated scheme from \citet{huang2015a}. Extinctions were calculated using the method by \citet{huber2017a} and \citet{berger2018a} and corrected  by adopting the extinction laws from \citet{yuan2013a}. Since \numax\ and amplitude cannot be measured for all the stars in our high-luminosity red-giant sample in the same way as for lower-luminosity stars, we used the frequency and height of the highest peak to represent \numax\ and the oscillation amplitude. We note that the group of Miras (green dots) have higher amplitudes than expected from the trend of the stars with higher \numax. This is because pulsations in Miras are driven differently from the rest of the sample. This seismic H-R diagram indicates that both \numax\ and the amplitude span more than six orders of magnitude, which represents so far the largest parameter ranges measured only from observations. 
Considering the availability of Gaia DR2 parallaxes for almost all of the stars shown in Figure~\ref{fig:hr}a, we plot a \mbox{H-R} diagram as shown in Figure~\ref{fig:hr}b, where luminosities were computed from Gaia DR2 parallaxes either from this work for our sample or from \citet{berger2018a} for the rest of the stars shown in Figure~\ref{fig:hr}b. We can see the red clump with a relatively large scatter. 

\redbf{It is challenging to distinguish RGB from AGB stars with similar effective temperatures and luminosities, although \citet{mosser2014a} have found a few dozens of stars leaving the main region of the red clump on their way to ascend the AGB. The tip of the RGB (TRGB) is a key feature in the H--R diagram where the density of star number starts to decrease significantly, allowing one to clearly identify stars in the AGB phase as those beyond the TRGB. This feature is, however, not clearly seen in Figure~\ref{fig:hr}b, which is caused by the limited sample size of stars near the TRGB and the large luminosity uncertainties. Although the \numax\ estimates are more precise, the TRGB is still not clearly seen in Figure~\ref{fig:hr}a.}

\section{Conclusions}
We carried out asteroseismic analyses of high-luminosity \kep\ red giants with pulsation periods P $\gtrsim$ 1 day.  We attempted to address open questions regarding the excitation mechanisms, radial order assignment, and dominant mode nature (radial or non-radial). We also investigated the relation between pulsation amplitude and period for low-\numax\ sun-like oscillators (\numax\ $\lesssim$ 10\muHz), SRs, and Mira variables. For the first time we performed a test on the validity of the seismic scaling relations with high-luminosity \kep\ red giants using Gaia DR2 parallaxes. The main results are summarized below:
\begin{enumerate}
    
    \item By comparing the amplitudes measured from full-length \kep\ and OGLE light curves and 1/3 shorter segments of the light curves, SRs are confirmed to be stochastically excited as solar-like oscillators, which is different from self-excited Mira variables. Using the same method, we find Mira variables have much longer mode lifetime than SRs, and the lifetime of SRs changes continuously (see Figure~\ref{fig:modelifttime}).
    
    \item We have made an unambiguous detection of well-resolved pulsation ridges, corresponding to radial fundamental mode and overtones ($2\leq n \leq7$), and sub-ridges, linked to $l=0,1,2$ modes (see Figure~\ref{fig:stackedspectra}). Our radial order assignment from the two ways (model frequencies and peak-bagging) is consistent with \citet{stello2014a}, and \citet{mosser2013a} for $n\geqslant3$ but not for $n=2$ (see Figure~\ref{fig:ridges} and \ref{fig:pattern}).
    
    \item Clear pulsation sequences on the P--L diagram have not been detected in \kep\ LPVs, which are expected to be present as for the OGLE LPVs in the LMC. The approximate six times larger uncertainty in absolute magnitude, $\sigma_{M_K}$, for \kep\ LPVs makes the ridges difficult to be detected (see Figure~\ref{fig:kperiod}).
    
    \item We show that the $l=1$ modes are dominant in the overtones of $n$=3, 4, 5, and 6, while the $l=2$ modes appear to be dominant in the first overtone $n=2$ (see Figure~\ref{fig:amplitude}). Since the triplet structure gets gradually closer with decreasing pulsation frequency, longer time series are required to resolve multiple angular degrees. OGLE light curves with a typical baseline of 8 years, with an extension to 12 years for some pulsators, are thus very valuable to resolve the first overtone and radial fundamental modes.
    
    \item A comparison of radii computed from the scaling relations to those derived from Gaia DR2 parallaxes shows good agreement, with an increasing systematic offset when \numax~$\lesssim$~3\muHz~(R~$\gtrsim$~40 R$_{\odot}$, or log $\rm L/L_{\odot}$~$\gtrsim$~2.6). This suggests the seismic scaling relations could break down in this regime. On the other hand, the comparison shows an excellent agreement where \numax~$\gtrsim$~3\muHz, implying that the scaling relations are still accurate. This also confirms the existence of the 0.03 mas systematic offset in Gaia DR2 parallaxes (see Figure~\ref{fig:radiicompnumax}).

\end{enumerate}

\tess\ will observe bright stars in nearly the whole sky and provide up to one year of time series for targets in the continuous viewing zones. These upcoming new data will permit the detection of known and potentially new LPVs, in particularly these with periods around 60 days, which has been argued as a threshold of substantial pulsation-triggered mass loss \citep{mcdonald2016a}. The supplementary ground-based spectroscopic observations and the high-quality parallaxes from Gaia and \textit{Hipparcos} enable us to investigate the relation between luminosity, metallicity, and pulsation, and hence shed light on the mechanism transition in mass loss from magneto-acoustic wind to pulsation wind \citep{mcdonald2018b}.

\section*{Acknowledgements}
We are thankful for the referee's careful reading of the manuscript and helpful comments and suggestions. We thank Jennifer van Saders, James S. Kuszlewicz, Nathalie Themessl, and Patrick Gaulme for their useful comments and discussions. We gratefully acknowledge the entire \kep\ team and everyone involved in the \kep\ mission for making this paper possible. Funding for the \kep\ Mission is provided by NASA\textquotesingle s Science Mission Directorate. This work was supported in part by the German space agency (Deutsches Zentrum f\"{u}r Luft- und Raumfahrt) under PLATO data grant 50OO1501. The computational resources were provided by the German Data Center for SDO through a grant from the German Aerospace Center (DLR). This work is partially supported by the Joint Research Fund in Astronomy (U1631236) under cooperative agreement between the National Natural Science Foundation of China (NSFC) and Chinese Academy of Sciences (CAS). D.H. acknowledges support by the National Science Foundation (AST-1717000). D.S. is the recipient of an Australian Research Council Future Fellowship (project number FT1400147). The research leading to the presented results has received funding from the European Research Council under the European Community's Seventh Framework Programme (FP7/2007-2013) / ERC grant agreement no 338251 (StellarAges).

\bibliographystyle{mnras}
\bibliography{reference.bib}
%%%%%%%%%%%%%%%%%%%%%%%%%%%%%%%%%%%%%%%%%%%%%%%%%%

%%%%%%%%%%%%%%%%%%%%%%%%%%%%%%%%%%%%%%%%%%%%%%%%%
% Don't change these lines
\bsp	% typesetting comment

\end{document}